\documentclass[longauth]{aa} 
\usepackage{amsmath}	
\usepackage{amssymb}	
\makeatletter
\renewcommand\@biblabel[1]{}
\makeatother
\usepackage{graphicx}
\usepackage{graphicx}
\usepackage{txfonts}
\usepackage{xcolor}

\begin{document}

   \title{Stellar tidal streams around nearby spiral galaxies with deep imaging from amateur telescopes}
\titlerunning{Deep imaging with amateur telescopes}
    \authorrunning{Mart{\'\i}nez-Delgado et al.}
    
   \author{David Mart{\'\i}nez-Delgado\inst{1,2,3}\thanks{ARAID Fellow}, Michael Stein\inst{4}, Joanna D. Sakowska\inst{5,3}, M. Maurice Weigelt \inst{4}, Javier Roman\inst{5,6}, Giuseppe Donatiello\inst{7}, Santi Roca-F\`abrega\inst{8}, Mischa Schirmer$^{9}$, Eva K. Grebel\inst{10}, Teymoor Saifollahi\inst{11}, Jeff Kanipe\inst{12}, M. Angeles G\'omez-Flechoso\inst{5,6}, Mohammad Akhlaghi\inst{1}, Behnam Javanmardi\inst{13}, Gang Wu\inst{14,15}, Sepideh Eskandarlou\inst{1}, Dominik J. Bomans\inst{4}, Cristian Henkel\inst{15},  Adam Block\inst{16}, Mark Hanson\inst{17}, Johannes Schedler\inst{18}, Karel Teuwen\inst{19}, R. Jay GaBany \inst{20}, Alvaro Iba\~nez Perez\inst{21}, Ken Crawford\inst{22},  Wolfgang Promper\inst{23}, Manuel Jimenez\inst{24}, S\'{i}lvia Farr\`{a}s-Aloy\inst{25}, Juan Mir\'{o}-Carretero\inst{5,6}}
   
   \institute{
$^{1}$ Centro de Estudios de F\'isica del Cosmos de Arag\'on (CEFCA), Unidad Asociada al CSIC, Plaza San Juan 1, 44001 Teruel, Spain\\
$^{2}$ ARAID Foundation, Avda. de Ranillas, 1-D, E-50018 Zaragoza, Spain\\
$^{3}$ Instituto de Astrofísica de Andalucía, CSIC, Glorieta de la Astronom\'\i a,  E-18080 Granada, Spain\\
$^{4}$ Ruhr University Bochum, Faculty of Physics and Astronomy, Astronomical Institute (AIRUB), 44780 Bochum, Germany \\
$^{5}$ Department of Physics, University of Surrey, Guildford GU2 7XH, UK \\
$^{5}$Departamento de F{\'\i}sica de la Tierra y Astrof{\'\i}sica, Universidad Complutense de Madrid, E-28040 Madrid, Spain\\
$^{6}$Instituto de F{\'\i}sica de Part{\'\i}culas y del Cosmos (IPARCOS), Universidad Complutense de Madrid, E-28040 Madrid, Spain\\
$^{7}$ UAI -- Unione Astrofili Italiani /P.I. Sezione Nazionale di Ricerca Profondo Cielo, 72024 Oria, Italy \\
$^{8}$ Lund Observatory, Division of Astrophysics, Department of Physics, Lund University, Box 43, SE-221 00 Lund, Sweden \\
$^{9}$ Max-Planck Institute for Astronomy, K\"onigstuhl 17, 69117, Heidelberg, Germany\\
$^{10}$ Astronomisches Rechen-Institut, Zentrum f{\"u}r Astronomie der Universit{\"a}t Heidelberg, M{\"o}nchhofstr.\ 12--14, D-69120 Heidelberg, Germany\\
$^{11}$  Université de Strasbourg, CNRS, Observatoire astronomique de Strasbourg (ObAS), UMR 7550, F-67000 Strasbourg, France\\
$^{12}$ Left Hand Observatory, Boulder, CO, USA\\
$^{13}$ Argelander Institut für Astronomie der Universität Bonn, Auf dem Hügel 71, 53121 Bonn, Germany\\
$^{14}$ Xinjiang Astronomical Observatory, Chinese Academy of Sciences, 830011 Urumqi, Xinjiang, PR China\\
$^{15}$ Max-Planck-Institut f{\"u}r Radioastronomie, Auf dem H{\"u}gel 69, 53121 Bonn, Germany \\
$^{16}$ Steward Observatory, Department of Astronomy, University of Arizona, 933 N. Cherry Avenue, Tucson, AZ 85748, USA\\
$^{17}$ Doc Greiner Research Observatory-Rancho Hidalgo, Animas, New Mexico, USA \\
$^{18}$ CHART32, CTIO, Chile\\
$^{19}$ Remote Observatories Southern Alpes, Verclause, France\\
$^{20}$ Black Bird Observatory II, Alder Springs, California, USA \\
$^{21}$ Asociación Astronómica AstroHenares, 28823 Coslada, Madrid, Spain\\
$^{22}$ Rancho del Sol Observatory, Camino, California, USA\\
$^{23}$ Tivoli Southern Sky Guest Farm, Namibia\\
$^{24}$ MJ Observatory, Cuenca, Spain\\
$^{25}$ Universidad Internacional de Valencia (VIU), C. del Pintor Sorolla 21, 46002 Valencia, Spain\\  
}

   \date{}

 
  \abstract
   {Tidal interactions between massive galaxies and their satellites are fundamental processes in a Universe with $\Lambda$-Cold Dark Matter ($\Lambda$CDM) cosmology, redistributing material into faint features that preserve records of past galactic interactions. While stellar streams in the Local Group impressively demonstrate satellite disruption, they do not constitute a statistically significant sample. Constructing a substantial catalog of stellar streams beyond the Local Group remains challenging due to the difficulties in obtaining sufficiently deep, wide-field images of galaxies. Despite their potential to illuminate dark matter distribution and galaxy formation processes overall, stellar streams remain underutilized as cosmological probes.}
   {The Stellar Tidal Stream Survey (STSS) addresses this observational gap by leveraging amateur telescopes to obtain deep, scientific-grade images of galactic outskirts, capable of building a more statistically meaningful sample of stellar streams.}
   {Over the last decade, the STSS has acquired deep (up to surface brightness limit $\sim$ 28.3 mag/arcsec$^2$ in the \textit{r}-band) wide-field images of 15 nearby Milky Way analog galaxies using a coordinated network of robotic amateur telescopes, avoiding the issues associated with `mosaicing’ smaller images taken with a single, professional telescope.}
   {Our survey has revealed a diverse range of previously unreported faint features related to dwarf satellite accretion— including stellar streams, shells, and umbrella-like structures. We serendipitously discover an ultra-diffuse galaxy (NGC150-UDG1) which shows hints of tidal tails.}
   {The STSS demonstrates the suitability of modern amateur telescopes to detect and study faint, diffuse structures in large fields around nearby spiral galaxies. Their economic and accessibility advantages enable larger statistical samples with deep imaging, essential for testing galaxy formation models and constraining the frequency and properties of minor merger events in the local Universe.}


\maketitle
%

\section{Introduction}

While the cosmological models built within the $\Lambda$-Cold Dark Matter ($\Lambda$CDM) paradigm predict a decline in minor merger rates to the present-day epoch (e.g., \citealt{Bullock2005, Cooper2010, Cooper2013}), they also suggest that satellite disruption remains common around most galaxies, particularly the more massive ones (\citealt{Guo2008, Jackson2022}). Consequently, galactic halos should contain various diffuse structural features resulting from interactions with dwarf satellites, globular clusters, and numerous dwarf-galaxy sized sub-halos (\citealt{Moore1999, Johnston2008}). The most spectacular cases include long, dynamically cold stellar streams which wrap around the host galaxy's disk, isolated shells, jet-like features, giant debris clouds, and large diffuse structures that are old, phase-mixed remnants of accreted satellites.

\begin{figure*}[t]

\includegraphics[width=1.0\textwidth]{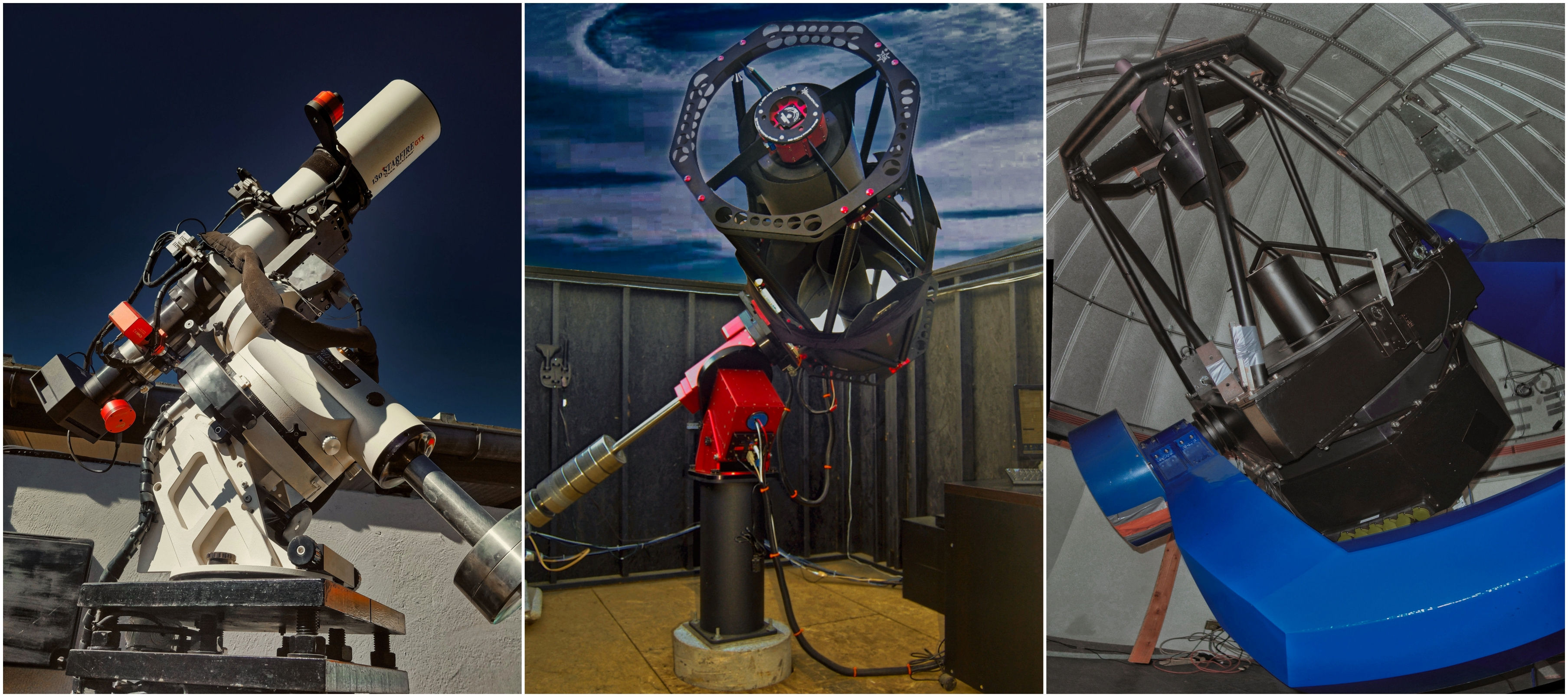}
    \includegraphics[width=1.0\textwidth]{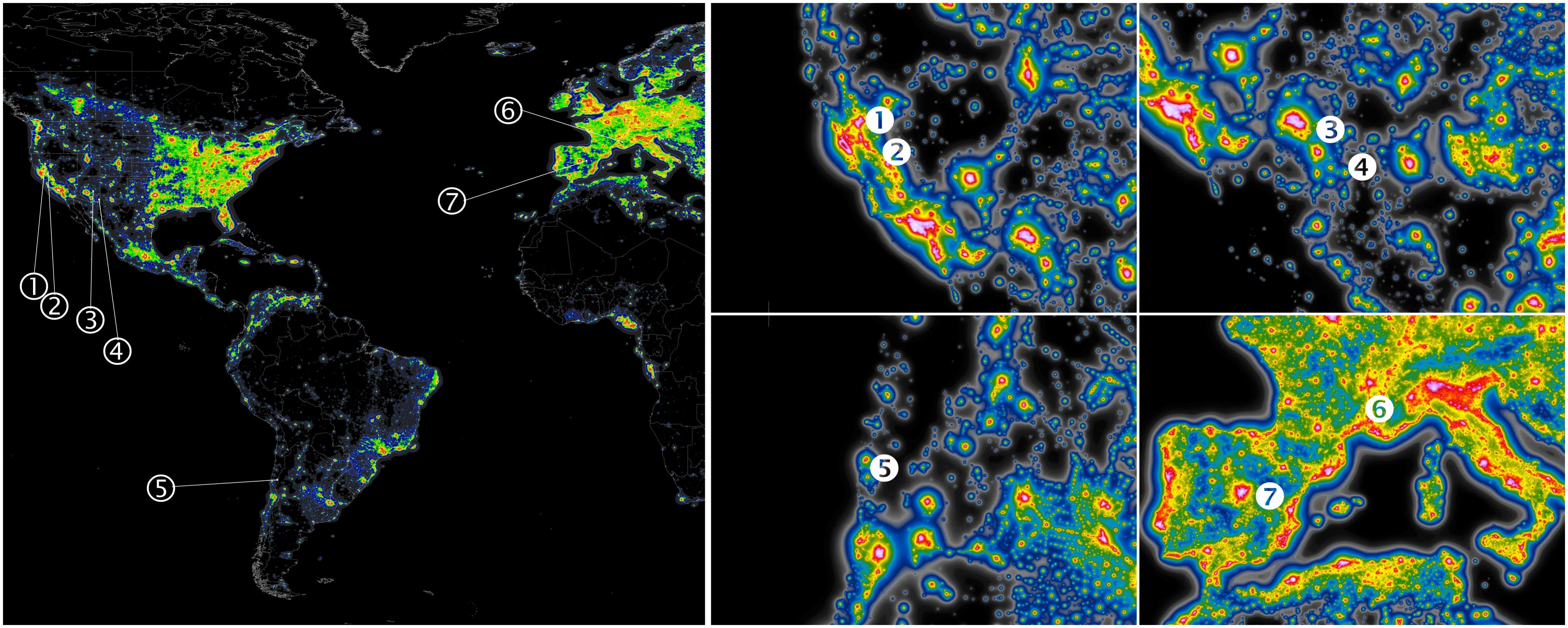}
    \caption{{\it Top row panels}: Three of the amateur telescopes used in the Stellar Tidal Stream Survey: (left) AstroPhysics 0.13-m  f/4.5; (center) RCO 0.506-m f/8.1; (right) CHART32 Cassegrain 0.80-m f/7. {\it Bottom row panels}: The location of the robotic telescope facilities used in the Stellar Tidal Stream Survey (see Table 1) are overplotted to the Light Pollution Map NITESat (\citealt{Falchi2016}): (1) Black Bird II Observatory; (2) Rancho del Sol Observatory (3) Mount Lemon Sky Center; (4) Rancho Hidalgo Observatory; (5)  CHART32; (6) ROSA Observatory; (7)  MJ Observatory. }\label{fig:telescopes}

\end{figure*}

Comparing the observed frequency and properties of stellar streams with simulations can verify whether $\Lambda$CDM correctly predicts the abundance and structure of tidal features (e.g., \citealt{Miro-Carretero2024a}), offering insights into baryonic processes within dark matter halos (e.g. \citealt{Shipp2023}). Beyond tracing merger histories, characterizing stellar streams around massive galaxies enables valuable tests of N-body simulations of tidal disruption and accretion. These models are constructed by fitting the sky-projected features from deep images, and dynamical analysis of these complex tidal structures offers unique constraints on the dark matter halos and their asymmetries. The primary challenge in modeling streams with imaging data alone is the degeneracy between orbit and inclination- fortunately even a few line-of-sight velocity measurements can help break this degeneracy, particularly for multi-wrap streams and measurements from opposite sides of the galaxy (e.g. \citealt{Walder2024}). 
Presently, obtaining radial velocities of hundreds of tidal debris stars with adequate signal-to-noise (S/N$\geqq$20) is challenging with ground-based facilities, although future surveys with, for instance, the Extremely Large Telescope (ELT, e.g. \citealt{ELT}) and MOSAIC (e.g. \citealt{mosaic}) offer significant promise.

Imaging stellar tidal streams around galaxies is inherently difficult; for nearby systems, their typical SB is \textit{at least} 26 mag/arcsec$^{2}$ or fainter, depending on the progenitor's luminosity and accretion time (\citealt{Johnston2001, Sola2025}). As such, streams around galaxies beyond the Local Group cannot be resolved into individual stars with modest telescopes but instead appear as elongated diffuse light regions extending over several arcminutes on the sky. Identifying LSB features around galaxies requires wide-field, deep images with excellent flat-field quality, covering extensive regions around target galaxies.


At the beginning of this century, only a few extragalactic stellar tidal streams had been reported in the local Universe. Using special contrast enhancement techniques on deep photographic plates, \cite{Malin1997} first highlighted the tidal features surrounding M83 and M104 (see \citealt{MD2021} for amateur images of M104). The detection of these faint extragalactic stellar streams, together with the simultaneous discovery of the Sagittarius tidal streams wrapping around the Milky Way (\citealt{Mateo1997, Ibata2001, MD2001, Majewski2003}), encouraged systematic searches for analogous tidal structures in nearby galactic halos. However, as professional telescopes leveraged the first wide-field cameras, they still suffered additional challenges when imaging extended LSB features. Their long focal ratios and large flat-field correction errors limited stellar stream discoveries to only a few cases (e.g., \citealt{Shang1998, Forbes2003, Pohlen2003}) for several years.

In this context, the Stellar Tidal Stream Survey (STSS)--an innovative professional--amateur (ProAm) collaboration involving astrophotographers worldwide operating robotic telescopes (see Fig. \ref{fig:telescopes})— demonstrated the advantages of small telescopes in detecting LSB features across large sky areas (\citealt{MD2008}). Small, short-focal-length telescopes with single-chip cameras provided larger fields of view (20-120 arcmin) enabling easier flat fielding of regions around galaxies in comparison to the multi-chip detector arrays used with professional telescopes during that decade. Observations using professional telescopes suffered from significant image background variations (e.g. flat-fielding issues for different chips, fringes, scattered light) due to stitching together multiple pointings to cover the full spatial extent of galactic halos ("mosaicing"). These artifacts complicated faint structure detection (see \citealt{Tal2009, Miskolczi2011}), and their correction added significant overhead to data gathering and processing. Finally, this project extended amateur contributions to a new research field relating to galaxy formation and evolution. The competitive nature of time allocation on professional telescopes makes it difficult to secure the substantial observing time needed for comprehensive surveys of galaxy outskirts, especially when the detection of faint features cannot be guaranteed in advance. 



Since its 2008 inception, the STSS has produced deep, wide-field images of nearby Milky Way-analogue galaxies in the Local Volume, revealing an assortment of large-scale tidal structures in the halos of select nearby galaxies consistent with those predicted by cosmological models (e.g., see Fig. 2 in \citealt{MD2010}). These include giant great circles, which are either intact or fragmented (NGC 4631/M104; \citealt{MD2015, MD2021}), giant umbrellas (NGC 922, \citealt{NGC922-2023}), shells (\citealt{Cooper2011}), thin "dog leg" streams (NGC 1097, \citealt{Amorisco2015}), loops and arcs (\citealt{Chonis2011}). Further, the STSS detected star formation caused by a minor merger (NGC 5387, \citealt{Beaton2014}) and within a stream (NGC 7241, \citealt{MD2024}). STSS also characterised low-mass galaxies, detecting the first stellar stream around a dwarf (NGC 4449, \citealt{MD2012}) and the tidal disruption of a dwarf spheroidal by its host beyond the Local Group for the first time (NGC 253-dw2, \citealt{Romanowsky2016}). The STSS additionally discovered new LSB dwarf satellites through a sister project DGSAT (Dwarf Galaxy Survey with Amateur Telescopes; \citealt{Javanmardi2016, Henkel2017}).  The use of amateur telescopes to image LSB features was further explored by other groups, such as the Dragonfly project  (\citealt{AbrahamVanDokkum2014, vanDokkum2014, Merritt2016}) an array of 2 x 24 Canon telephoto camera lenses, and the HERON survey (\citealt{Rich2019}).\\


In this paper, we present a compilation of the most remarkable results from the first decade of the STSS. In Section \ref{sec:observations}, we overview our observational strategy and introduce other data sets with which we leverage our analysis. In Section \ref{sec:Photometry}, we discuss our photometry procedure. Finally in Section \ref{sec:Results} we present our results and, in Section \ref{sec:Conclusions}, discuss the future of amateur astrophotography and our conclusions.

 \begin{center}
  \begin{table*}[hbt!]
      \caption[]{List of observatories and equipment}
         \begin{tabular}{l l l l l l l}
            \hline
            \noalign{\smallskip}
            Observatory & Location & SQM & Telescope & CCD camera & pixel scale \\
            \noalign{\smallskip}
            \hline
            \noalign{\smallskip}
Black Bird II (BBO II)      &        New Mexico (USA)     & 21.9   &                 RCO 0.506-m f/8.1   &       Apogee Alta U16M    &    0.46”/pix \\
Rancho del Sol (RdS)  &    California (USA)   &  21.6    &                      RCO 0.506-m f/8.1    &       Alta KAF 0900     &  0.58”/pix \\
Mount Lemmon Sky Center  &    Arizona (USA)     & 21.9    &           Schulman 0.80-m f/7     &         SBIG STX16803   &  0.33”/pix \\
CHART32        &                       CTIO, Chile &  21.9   &                              Cassegrain 0.80-m f/7   &           FLI PL-16803   &     0.33”/pix \\
ROSA       &                               Verclause, France & 21.7   &                     0.40-cm f/3.75    &      FLI ML-16803       &       1.24”/pix\\
Rancho Hidalgo Obs. (RHO)  &  New Mexico, USA   &  21.9  &                   RCSO 0.362-m  f/7.9    &       Apogee U16M  &    0.62”/pix \\
MJ Observatory (MJO)   &                   Huete, Spain   & 21.7  &                          AstroPhysics 0.13-m  f/4.5   & Moravian G3 11002 &    3.13 "/pix     \\
            \noalign{\smallskip}
            \hline
         \end{tabular}
   \end{table*}\label{galaxies}
 \end{center}

\section{Observations and data reduction}
\label{sec:observations}

\subsection{Stellar Tidal Stream Survey}
The observations of the STSS were conducted with seven privately owned observatories (located in Europe, the United States, and Chile) equipped with modest-sized telescopes (0.1-0.8 meters) utilizing the latest-generation commercial astronomical CCD camera. Each observing location features spectacularly dark, clear skies with seeing below 1.5''. The survey strategy strives for multiple deep exposures of each target using high-throughput clear filters with near-IR cut-off, known as luminance (L) filters (4000 \AA $< \lambda <$7000 \AA) and a typical exposure times of 7-8 hours. Our typical 3-$\sigma$ SB detection limit (measured in random apertures of 2" diameter) is $\sim$ 28 and 27.5 mag/arcsec$^2$ in $g$ and $r$, respectively, which is approximately two magnitudes deeper than the Sloan Digital Sky Survey (SDSS, \citealt{york2000}) DR8 images. In \ref{appendix} we provide recommendations for obtaining good images with amateur telescopes.




\subsection{Neutral hydrogen data}

To check for a link between NGC\,925 and the LSB overdensity observed within its image, we use the first data release (DR1) of the Hydrogen Accretion in LOcal GAlaxieS (HALOGAS) Survey (\citealt{Heald2011}) to trace the distribution of neutral hydrogen (H$\textsc{i}$) between the two objects. The data were observed using the Westerbork Synthesis Radio Telescope (WSRT) in its Maxishort configuration to optimize the imaging performance for extended targets.
The correlation backend was set up to provide two linear polarizations in 1024 channels with a 10\,MHz bandwidth centered at the systemic velocity of NGC\,925 (-16\,${\rm km\,s}^{-1}$).
Offline Hanning smoothing was used to lead to a final velocity resolution of about 4 \,${\rm km\,s}^{-1}$. To recover most of the diffuse gas, we used the low-resolution (LR) datasets. The LR data were imaged with a robust parameter of 0 within the miriad's ``invert" task \citep{Sault1995} and also an addition Gaussian $u, v$ taper corresponding to 30$^{\prime\prime}$ in the image plane, resulting in a synthesized beam of 37$\,.\!\!^{\prime\prime}$9 $\times$ 33$\,.\!\!^{\prime\prime}$2
with position angle (PA)$\approx$0$\,.\!\!^{\circ}$6 and a 1-$\sigma$ noise level of 0.17 Jy\,beam$^{-1}$ in a single channel.
We refer to \cite{Heald2011} for detailed information on the observations.

\subsection{DESI Legacy Imaging Survey}
For comparison purposes and for the photometry analysis of the NGC 150-UDG1 (see Section 4), we have also used image cutouts from the public, deep imaging data released recently by the DESI Legacy Imaging Survey (DESI LS; \citealt{Dey2019}) in three optical bands ($g$,$r$, and $z$), as described in \cite{MD2023SSLS}. For this paper, we have used data from the Dark Energy survey (DES; \citealt{DES2016}) which covers 5000 deg$^{2}$ of the southern sky using the Dark Energy Cam (3 deg$^{2}$ field of view, FOV; \citealt{Flaugher2015}) installed on the Blanco 4-m telescope at the Cerro Tololo Interamerican Observatory in Chile. The average SB limit of these images in the $r$-band is 28.65 mag/arcsec$^{2}$ (see Section 2.3 of \citealt{Miro-Carretero2024b}).

 \begin{center}
  \begin{table*}[t]
  \centering
      \caption[]{List of targets.}
         \begin{tabular}{l l l l l l l c  }
            \hline
            \noalign{\smallskip}
            Galaxy & RA (J2000) & Dec (J2000) & $D$ &Facility & Date & Exp. Time &  Field of View \\
                    &[H:M:S] & [D:M:S] & [Mpc] & & &[min]\\ 
            \noalign{\smallskip}
            \hline
            \noalign{\smallskip}

NGC~95  & 00h22m13.54s & +10d29m30.0s & 59.7\textsuperscript{TF} & RHO  &  Oct 2018&  600m  & 20$\arcmin$ $\times$ 20 $\arcmin$ \\
NGC~150  & 00h34m15.48s & -27d48m12.9s & 20.9\textsuperscript{SN} & CHART32   & Sept 2016 &   480m  & 15$\arcmin$ $\times$ 15 $\arcmin$ \\
ESO~545-5 &02h20m06.11s & -19d45m02.6s & 30.1\textsuperscript{TF} & CHART32   & Aug 2017   &600m  & 7$\arcmin$ $\times$ 7 $\arcmin$\\
NGC~925  & 02h27m16.88s & +33d34m45.0s & 8.8\textsuperscript{C}  &   ROSA      &      Nov-Dec 2014  &    420m  & 30$\arcmin$ $\times$ 30 $\arcmin$\\
NGC~1511 & 03h59m36.98s & -67d38m03.3s & 14.2\textsuperscript{TF} & CHART32  & Sept-Nov 2014  &   400m  & 15$\arcmin$ $\times$ 15 $\arcmin$\\
NGC~2460 & 07h56m52.29s & +60d20m57.8s & 35.2\textsuperscript{TF}&   BBO II    &      Mar 2013      &    480m  & 15$\arcmin$ $\times$ 12 $\arcmin$\\
NGC~2775 & 09h10m20.12s & +07d02m16.6s & 17.0\textsuperscript{TF} &   MtLemmon   &     Jan-March 2011 &   200m  & 30$\arcmin$ $\times$ 20 $\arcmin$\\
NGC~3041 & 09h53m07.14s & +16d40m39.6s & 24.7\textsuperscript{TF} & RHO   &    Feb 2015    &    600m  & 20$\arcmin$ $\times$ 20 $\arcmin$\\
NGC~3614 & 11h18m21.32s & +45d44m53.6s & 34.7\textsuperscript{TF}&   MtLemmon    &    Febr 2015   &   720m   & 20$\arcmin$ $\times$ 20 $\arcmin$\\
NGC~3631 & 11h21m02.87s & +53d10m10.5s & 8.7\textsuperscript{TF} &   ROSA    &        Feb-Mar 2014 &     720m  & 20$\arcmin$ $\times$ 20 $\arcmin$\\
NGC~4390  & 12h25m50.71s & +10d27m32.2s & 29.21\textsuperscript{TF} &  MtLemmon   & April 2015 &  480m      & 10$\arcmin$ $\times$ 10 $\arcmin$\\
NGC~4414 & 12h26m27.10s & +31d13m24.7s & 21.4\textsuperscript{C} &   MtLemmon  &      Jan 2015     &     720m  & 30$\arcmin$ $\times$ 30 $\arcmin$\\
NGC~4684 & 12h47m17.52s	& -02d43m38.7s & 13.9\textsuperscript{SBF}&   RdS      &       Jan 2011   &       420m  & 15$\arcmin$ $\times$ 151 $\arcmin$\\
NGC~4826 & 12h56m43.67s & +21d40m58.7s & 7.5\textsuperscript{SBF} & MJO  & May-June 2019 &  350m    & 100$\arcmin$ $\times$ 100 $\arcmin$ \\
NGC~5750 & 14h46m11.12s & -00d13m22.6s & 32.4\textsuperscript{TF}& RHO   &         Mar 2015   &       600m   & 12$\arcmin$ $\times$ 12 $\arcmin$ \\    
NGC~5866 & 15h06m29.50s & +55d45m47.6s & 14.7\textsuperscript{SBF}& MtLemmon  &      Apr-May 2017   &   600m  & 20$\arcmin$ $\times$ 20 $\arcmin$ \\
NGC~7742 & 23h44m15.73s & +10d46m01.5s & 22.2\textsuperscript{TF} & RHO   &   Oct 2017    &    600m  & 18$\arcmin$ $\times$ 18 $\arcmin$\\



            \noalign{\smallskip}
            \hline
         \end{tabular}
         \tablefoot{ Coordinates and distances are taken from NASA/IPAC Extragalactic Database. For multiple available distance estimates, we rely on the most recent measurement for a given method. Here we prioritize the different distance estimation methods as follows: Cepheids (C), surface brightness fluctuations (SBF), supernovae (SNe), and Tully-Fisher (TF). }
   \end{table*}\label{targets}
 \end{center}

\section{Photometry}
\label{sec:Photometry}

In this Section, we describe the technique to calibrate the STSS luminance images to $r$-band magnitudes of the Panoramic Survey Telescope and Rapid Response System survey \citep[Pan-STARRS][]{2016arXiv161205560C} or SkyMapper survey \citep{2018PASA...35...10W,2019PASA...36...33O}, the process of estimating the SB limits of each field as well as other photometric measurements. We selected the $r$-band given it has the most wavelength overlap with our images and our former successful experience with previous STSS publications (see, e.g., \citealt{MD2015})\footnote{In fig. 5 of \citealt{MD2015} we additionally found that the main contribution to the light observed from the stellar stream is from old, metal poor RGB stars, strengthening our choice of the $r$-band}. First of all, world-coordinate system (WCS) calibration was performed using \texttt{astrometry.net}.\footnote{\url{http://astrometry.net/}}. As the typical FOV spans several arcminutes, there were always enough stars in the field to find an accurate WCS calibration with sub-arcsecond accuracy. 

\subsection{Photometric calibration}

We calibrated the deep luminance-filter images to the Pan-STARRS survey given most of our targets were also covered by Pan-STARRS. Where data was unavailable (ESO~545-5, NGC~150, and NGC~1511), we used catalog data from the SkyMapper Southern Sky Survey \citep{2018PASA...35...10W,2019PASA...36...33O}. Pan-STARRS generally follows the SDSS filter design- the Pan-STARRS $r_{\mathrm{P1}}$- and $i_{\mathrm{P1}}$-filter are very comparable to the SDSS $r$- and $i$-filter (\citealt{2012ApJ...750...99T,2016arXiv161205560C}, see \citealt{2010AJ....139.1628D} for a description of the SDSS r-band). The SkyMapper survey also presents an $r$-filter equivalent. In Table \ref{tab:filter_info} we list the relevant characteristics for the $r$-filter equivalent across all three surveys, demonstrating that these filters are very comparable and thus our calibration permits a comparison of our results to other SDSS-filter based surveys.

\begin{table}
    \centering
    \caption{Filter information of wide-field surveys that are used for the photometric calibration in comparison to the SDSS $r$-filter.}
    \begin{tabular}{lrrrr}
    \hline \hline
    Survey & $\lambda_{\mathrm{eff}}$ & $\lambda_{\mathrm{min}}$ & $\lambda_{\mathrm{max}}$ & W\textsubscript{eff} \\
           &[\AA]                     & [\AA]                      &   [\AA]                     & [\AA] \\
           \hline
    SDSS $r$                    & 6141 & 5415 & 6989 & 1056 \\
    Pan-STARRS r$_{\mathrm{P1}}$  & 6156 & 5386 & 7036 & 1252	\\
    SkyMapper $r$               & 6077 & 4925 & 7232 & 1414 \\
    \hline 
    \end{tabular}\label{tab:filter_info}
    \tablefoot{ We list the effective, minimal, and maximal wavelength as well as the effective width of each filter. Filter information was taken from the SVO-Filter Service \citep{2012ivoa.rept.1015R,2020sea..confE.182R}. }
\end{table}

\begin{figure}
   \centering
       \includegraphics[width=0.8 \columnwidth]{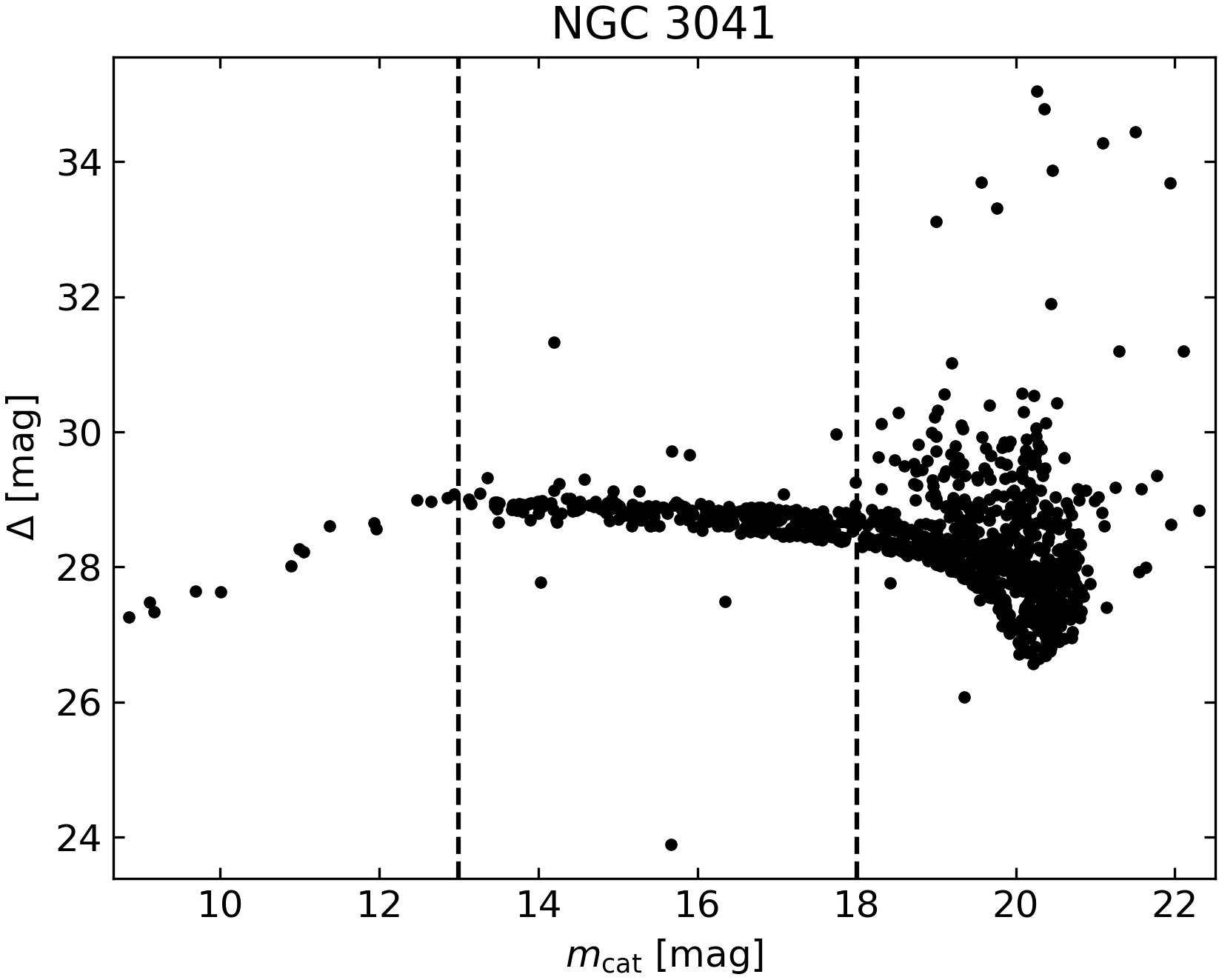}
   \caption{Calibration plot of NGC 3041. We show the difference of cross-matched STSS magnitudes and calibration magnitudes from the calibration catalog. The sources that lie outside of the dashed line were excluded from the calibration.}\label{fig:mag_zeropoint_example}
\end{figure}

\begin{figure}
  \centering
       \includegraphics[width=0.8\columnwidth]{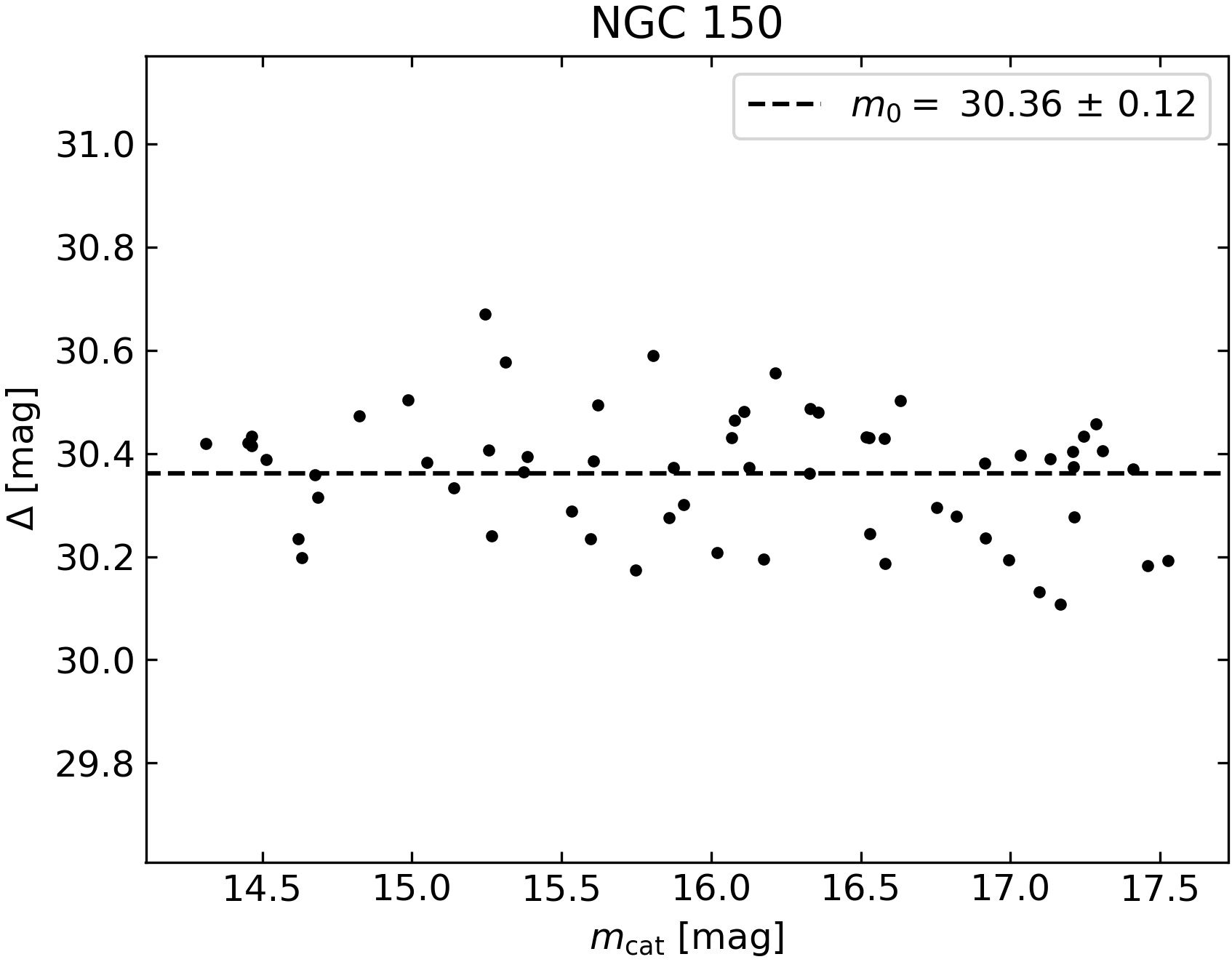}
   \caption{Zeropoint calibration for NGC~150. Zoom on the magnitude range that is used for the calibration process.}\label{fig:mag_zeropoint_zoom}
\end{figure}

We estimated the background value in each field with the SExtractor background algorithm \citep{1996A&AS..117..393B} that is implemented in the \texttt{photutils} Python package. In this step of the analysis, we only measured bright stars in the field and thus a single-value background estimation was sufficient. We then masked out all sources using a 3$\sigma$ detection threshold where a minimum of five connected pixels must lie above the threshold. For the detected source, we incorporated a dilation size of $10\times10$ pixels. Then, we computed and subtracted the global background estimate. To minimize the influence of photometric uncertainties on the calibration process, we only detected bright sources in our calibration source catalog by applying a 10$\sigma$ detection threshold. We cross-matched these sources accordingly with either the Pan-STARRS photometric catalog (DR1) or the SkyMapper catalog. We then derived an empirical zero-point by comparing the offset of the catalog magnitudes and the STSS measurements where we applied an arbitrary zero-point. In Fig. \ref{fig:mag_zeropoint_example}, we show the calibration plot of NGC~3041 as an example. Here, we can observe the typical characteristics when cross-matching two surveys. The plot is divided into three sections. In the left part, one can see deviations from very bright sources. 
The middle section generally shows good agreement between the STSS and the catalog magnitudes. As an example, we show a zoom on the calibration range for NGC~150 in Fig. \ref{fig:mag_zeropoint_zoom}. This section is well suited to compute the zero-point of our observations. The scatter in this section of the plot gives a measure of the uncertainty of our calibration process. There is a slight slope visible in the cross-matched sources, which might be caused slight non-linear effects of the used CCDs. Nevertheless, the slope is much smaller compared to the overall scatter of the population, and therefore, the $r$-filter can still be used as an anchor point for our observations. For all fields, we define the middle region to lie between m\textsubscript{max}=14\,mag and m\textsubscript{min}=18\,mag in the calibration catalog. In the right section of the plot (m>18, Fig. \ref{fig:mag_zeropoint_example}), deviations rise again as the catalog reaches its detection limit.

\subsection{Detection Limits}
\label{sec:detect_lim}


\begin{table*}
\caption{Detection limits for STSS galaxies derived with NoiseChisel. }
\label{tab:detect_lim}
    \centering
\begin{tabular}{l r r r }
    \hline\hline 
         Galaxy & ZP\textsubscript{r-filter} & $\mu_{\mathrm{lim}}^{\mathrm{aperture}}$   & $\mu_{\mathrm{stream}}$\\
                & [mag]  & [mag\,arcsec\textsuperscript{-2}]  & [mag\,arcsec\textsuperscript{-2}]\\
         \hline
NGC~95 & 29.49$\pm$0.13 & 27.10$\pm$0.13 &  26.61$\pm$0.14\\
NGC~150 & 30.36$\pm$0.12 & 26.31$\pm$0.12 &  25.06$\pm$0.12\\
        &  & (27.17$\pm$0.12) & \\
ESO~545-5 & 30.37$\pm$0.13 & 25.45$\pm$0.13  & 24.35$\pm$0.13\\
        &  & (26.99$\pm$0.13) &  \\
NGC~925 & 27.62$\pm$0.08 & 25.40$\pm$0.08 &  25.98$\pm$0.09\\
NGC~1511 & 29.81$\pm$0.10 & 24.54$\pm$0.10 &  24.40$\pm$0.10\\
        &  & (27.05$\pm$0.10)   & \\
NGC~2460 & 29.62$\pm$0.05 & 27.71$\pm$0.05 & 25.04$\pm$0.05\\
NGC~2775 & 30.34$\pm$0.15 & 25.86$\pm$0.15 & 25.03$\pm$0.15\\
        &  & (27.19$\pm$0.15) &  \\
NGC~3041 & 29.73$\pm$0.16 & 28.33$\pm$0.16 &  26.94$\pm$0.18\\
NGC~3614 & 31.14$\pm$0.16 & 27.39$\pm$0.16 &  27.25$\pm$0.16\\
NGC~3631 & 27.45$\pm$0.17 & 25.02$\pm$0.17 &  27.12$\pm$0.20\\
NGC~4390 & 30.63$\pm$0.17 & 26.82$\pm$0.17 &  25.54$\pm$0.17\\
NGC~4414 & 30.70$\pm$0.17 & 26.99$\pm$0.17 &  26.20$\pm$0.17\\
NGC~4684 & 29.84$\pm$0.14 & 26.97$\pm$0.14 &  26.31$\pm$0.14\\
NGC~4826 & 14.24$\pm$0.17 & 26.34$\pm$0.17 &  -\\
NGC~5750 & 28.86$\pm$0.16 & 26.98$\pm$0.16 &  25.03$\pm$0.16\\
NGC~5866 & 30.60$\pm$0.16 & 26.74$\pm$0.16 & 25.10$\pm$0.16\\
NGC~7742 & 18.19$\pm$0.14 & 26.85$\pm$0.14 &  25.83$\pm$0.14\\
\hline
    \end{tabular}
\tablefoot{Detection limits are computed for circular apertures with a size of 100\,arcsec\textsuperscript{2}. The upper-limit SB measurement based on placing apertures is indicated as $\mu_{\mathrm{lim}}^{\mathrm{aperture}}$. For galaxies with significant variation in the sky-$\sigma$ image, we list the limit as derived from the whole field versus cropped to a region of low sky-$\sigma$ (the latter which is stated in brackets). The uncertainties of the detection limits come from the calibration process from luminance to $r$-filter. Calibration errors and uncertainties from background noise (based on the aperture measurements) are propagated for the measurement of the stream SB $\mu_{\mathrm{stream}}$.}
\end{table*}

With the properly calibrated STSS images, we can now derive our detection limits (also known as the upper limit SB, see definition in \citealt{Miro-Carretero2024b}) for each field. To do this, we use the \texttt{NoiseChisel} \citep{2015ApJS..220....1A,Akhlaghi2019} task of the \texttt{GNU Astronomy Utilities}\footnote{\url{https://www.gnu.org/software/gnuastro/}}. To separate background regions from sources, we use a tesselation box size of $30\times30$ pixels and a narrow convolutional kernel with a full-width half maximum (FWHM) of two pixels.
We report the results of upper-limit SB measurements for circular apertures with a size of 100\, arcsec\textsuperscript{2}. In Table \ref{tab:detect_lim} we present the calibrated zero points, detection limits, and SB measurements of detected streams for all STSS galaxies. We note that some of the galaxies (NGC 150, ESO 545-5, NGC 1511, NGC 2775) show a strong variation in the sky standard deviation, which might point to problems in the flat fielding approach. For these galaxies, we report two SB limits (complete field vs. cropped to an area of low background deviation).

We conclude that we can reliably trace LSB structures in all analyzed datasets.
\subsection{Comparison with other surveys}
\begin{figure*}[hbt!]
    \centering
    \includegraphics[width=1.\linewidth]{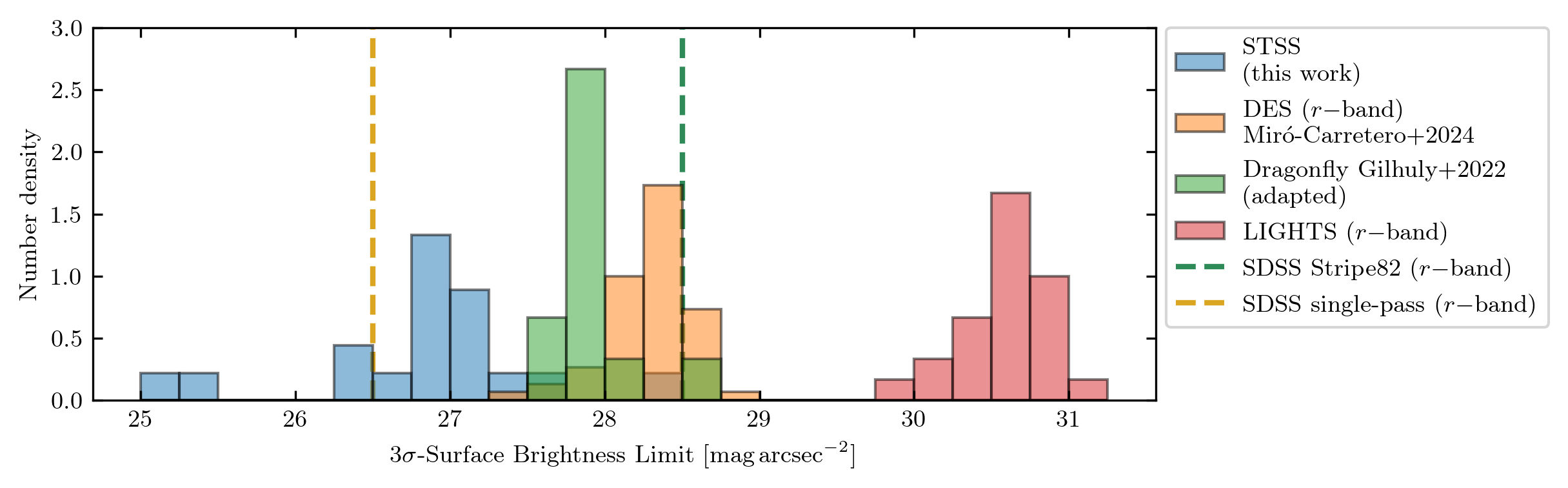}
    \caption{Surface brightness limits for different surveys, computed in areas equivalent to $10\times10$\,arcsec\textsuperscript{2} boxes: This work (STSS), Stellar Stream Legacy Survey (SSLS POC; proof of concept study using DES data, \citealt{Miro-Carretero2024b}), Dragonfly survey (\citet{Gilhuly2022}), LIGHTS survey $r-$band (\citealt{Zaritsky2024}), Stripe 82 and SDSS $r$-band (both taken from \citealt{fliri2016}).}\label{fig:sblimit_hist}
\end{figure*}

\begin{figure*}
\centering
	\includegraphics[width=1.0\textwidth]{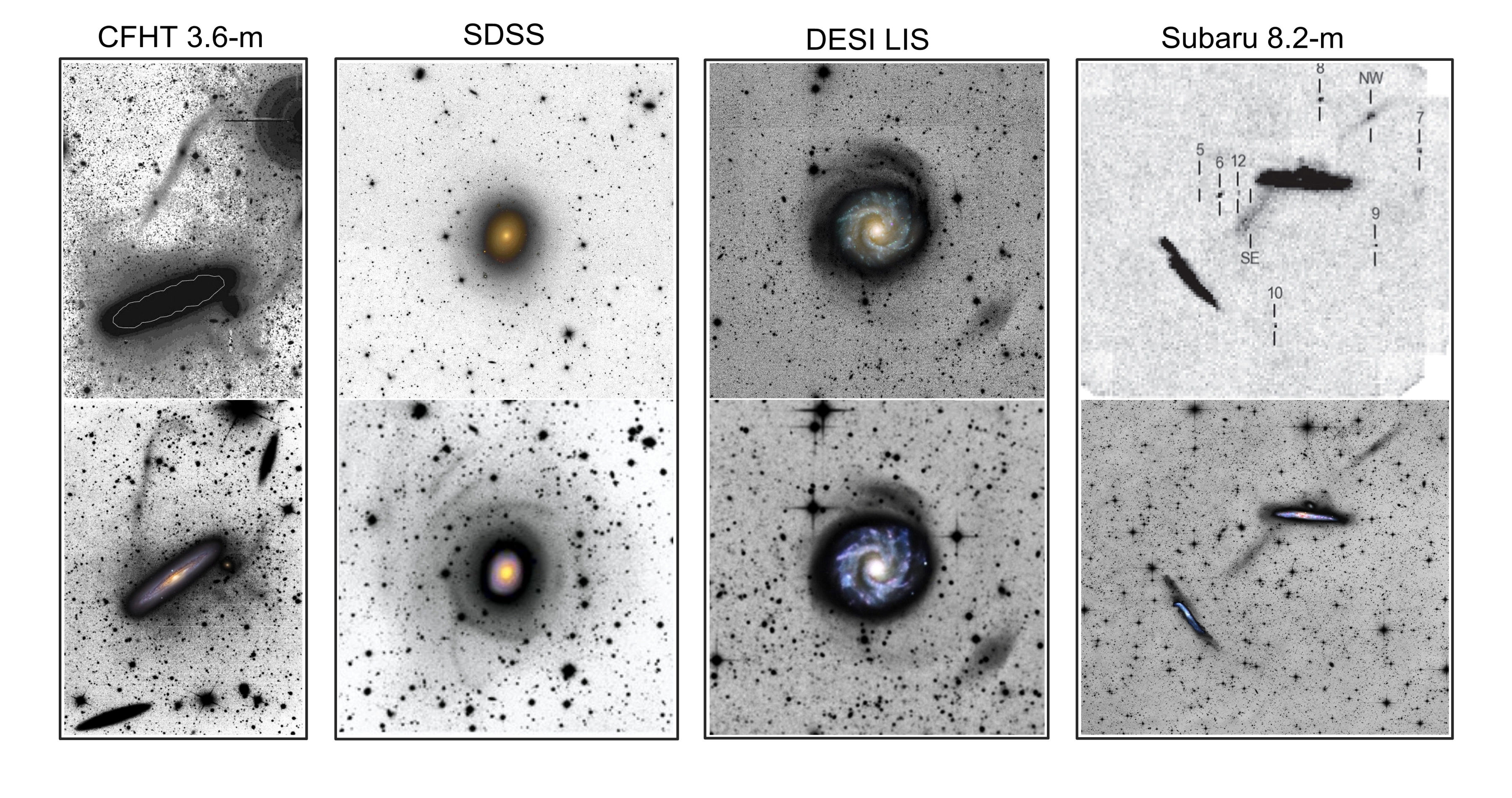}
    \caption{A comparison of deep amateur images from the STSS with data obtained with large professional telescopes (top) and large-scale CCD imaging surveys (bottom). From left to right: 
    ({\it top panels}): NGC 4216 taken with the Canada-French-Hawaii 3.6-m telescope (\citealt{Paude2013}); NGC 2275 from stacked images from the Sloan Digital Sky Survey (see \citealt{Morales2018}); NGC 3631 with the CTIO Blanco 4-m telescope (\citealt{MD2023SSLS}); and a stellar density map from the resolved red giant-branch stars of the stellar halo of NGC 4631 taken with the Subaru 8.2-m telescope (\citealt{Tanaka2017});
    ({\it bottom panels}): NGC 4216 taken with the BBO 0.5-m telescope (\citealt{MD2010}); NGC 2275 (this work; see Table 2); NGC 3631 (this work; see Table 2); and NGC 4631 obtained with the ROSA 0.3-m telescope \citealt{MD2015}). }\label{fig:comparison}
\end{figure*}

To better assess the photometric quality of our dataset, we compare it with other deep surveys in Fig \ref{fig:sblimit_hist}. Here, we compare the fields analyzed in this study (STSS) with DES data \citep{Miro-Carretero2024a} as well as single SDSS $r'$-band exposures and the SDSS stack from the IAC Stripe 82 Legacy Project \citep{fliri2016}. We also include data from the Dragonfly Edge-on galaxies survey \citep{Gilhuly2022} and the LBT Imaging of Galactic Halos and Tidal Structures (LIGHTS) survey \citep{Zaritsky2024}. \citet{Gilhuly2022} report SB limits on a scale of one arcmin\textsuperscript{2}, which makes a comparison with other data sets trickier. We therefore reevaluate the Dragonfly Edge-on galaxies survey data set with NoiseChisel\footnote{As the background noise increases towards the edge of the published data sets, we limit our surface brightness limit estimates to the central 2000 $\times$ 2000 pixels.} to apply the same metric as used for the STSS data and display the adapted SB limits in Fig. \ref{fig:sblimit_hist}. The distribution of the detection limits indicates that data sets obtained with amateur telescopes already enhance the ability to detect LSB structures compared to large-area surveys like the SDSS. The deepest images even reach a similar depth as those from the DES, highlighting the capabilities of amateur telescopes when proper calibration techniques and observation strategies are used.

\begin{figure*}
\centering
	\includegraphics[width=0.9\textwidth]{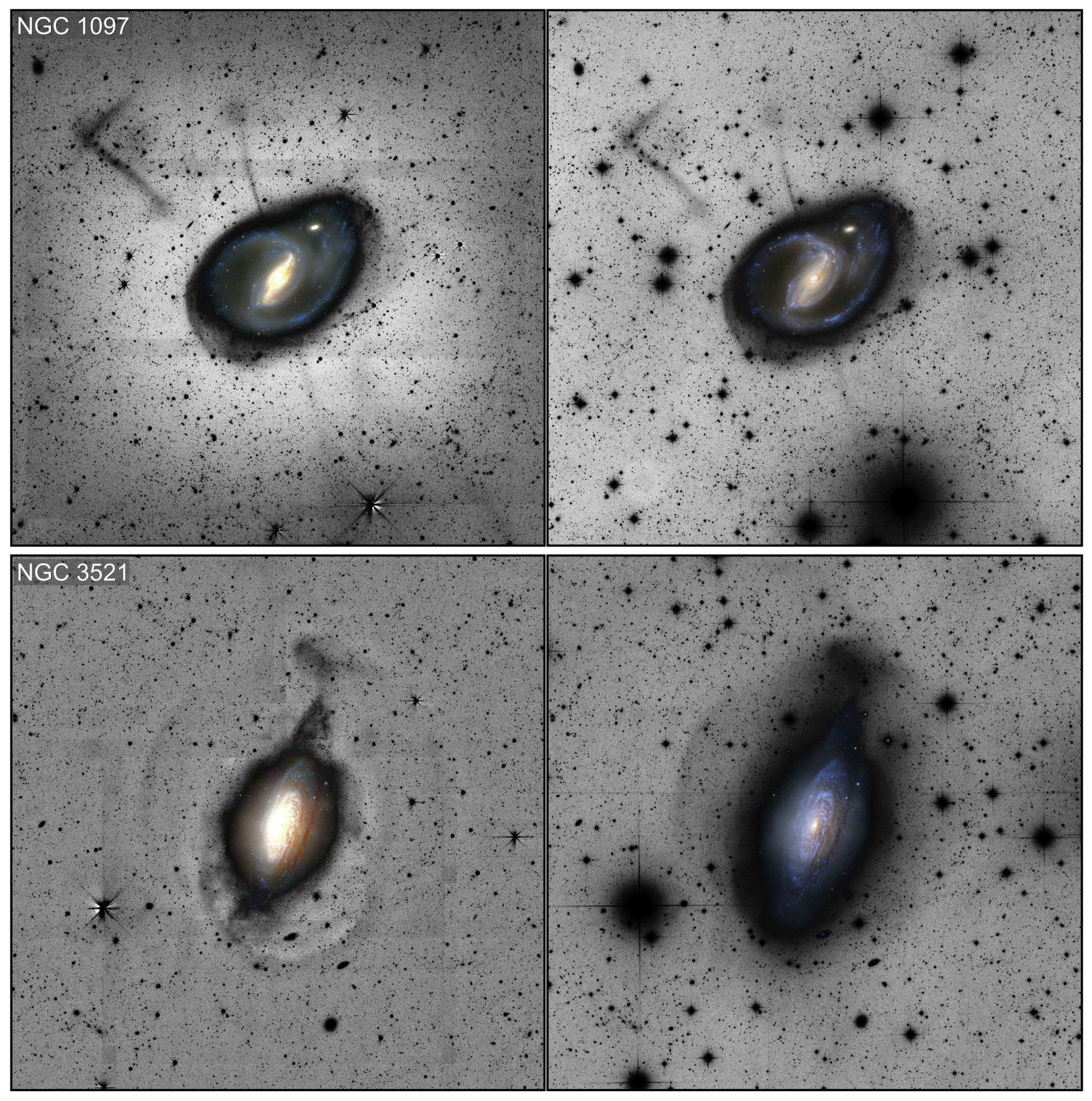}
    \caption{A comparison of DESI LS image cutouts computed with the {\it legacypipe} following the approach described in \citealt{MD2023SSLS} for NGC 1097 and NGC 3521 ({\it left panel}) versus deep amateur images obtained with a ASA600 RC 0.60-m f/f4.5 telescope ({\it right panel}).  The 3-$\sigma$ SB limiting magnitudes of these images are 27.87 and 28.07 mag/arcsec$^2$ for NGC 1097 and NGC 3521, respectively. The total FOV of both images is 30$\arcmin$ $\times$ 30 $\arcmin$.  It illustrates the advantages of our single-chip approach used in small telescopes for wide-field cameras versus large telescope data affected by the mosaicing problems described in Sec. 3.3.}\label{fig:mosaic}
\end{figure*}

Figure \ref{fig:comparison} shows that the results from the STSS are in excellent agreement with the deep images and stellar density maps obtained with professional telescopes. 
Here, we highlight that professional surveys tend to suffer from widely known problems of over-subtraction around objects of large apparent size (i.e., galaxies) during the reduction process. These problems are related to the flux adjustment in the outer parts of these structures in the sky subtraction procedure, thus eliminating the LSB information contained in the outer parts of galaxies. This problem is particularly important when the extent of a galaxy is either of the order of the size of an individual CCD in the instrumentation, or when it exceeds the typical grid size in the sky subtraction procedures. As a result, the LSB signal around bright galaxies is subtracted by default in wide-area surveys. We show some examples of this problem in Fig. \ref{fig:mosaic} for the case of DESI data. As for the Hyper Suprime-Cam Subaru Strategic Program (HSC-SSP), the sky over subtraction was found and fixed in HSC-SSP DR2 (see Figure 5 of \citealt{Aihara2019}), but re-introduced in DR3 intentionally (see Figure 8 of \citealt{Aihara2022}). This was because of the detection and segmentation algorithms they used which would loose significant completeness when objects become too connected. LSB-friendly detection and segmentation methods (like those of \citealt{Akhlaghi2019}) are immune to this problem.

While there are ongoing efforts to address the possible systematic effects of sky subtraction for galaxies or sources of large spatial extent for the Vera Rubin Observatory \citep{2024MNRAS.528.4289W}, and there are custom pipelines for Euclid that manage to preserve all LSB information even for very large objects \citep{2024arXiv240513496C, 2024arXiv240513499H}, these over-subtraction effects will be present in the default data releases due to the problems described above in their detection and segmentation algorithms. A custom reduction pipeline's development and execution (on such large datasets) would be very time consuming and require major resources.

Due to the large field-of-view of monolithic cameras belonging to amateur telescopes, this over-subtraction effect is easier to deal with than in surveys such as DESI, Euclid or LSST. Therefore, deep imaging from low-cost amateur telescopes provide an alternative way to stream discoveries in the local Universe while data processing techniques are improved sufficiently to deal with sky subtractions at all spatial scales.

\subsection{Galactic cirrus contamination}
\begin{figure*}
\centering
	\includegraphics[width=0.98\textwidth]{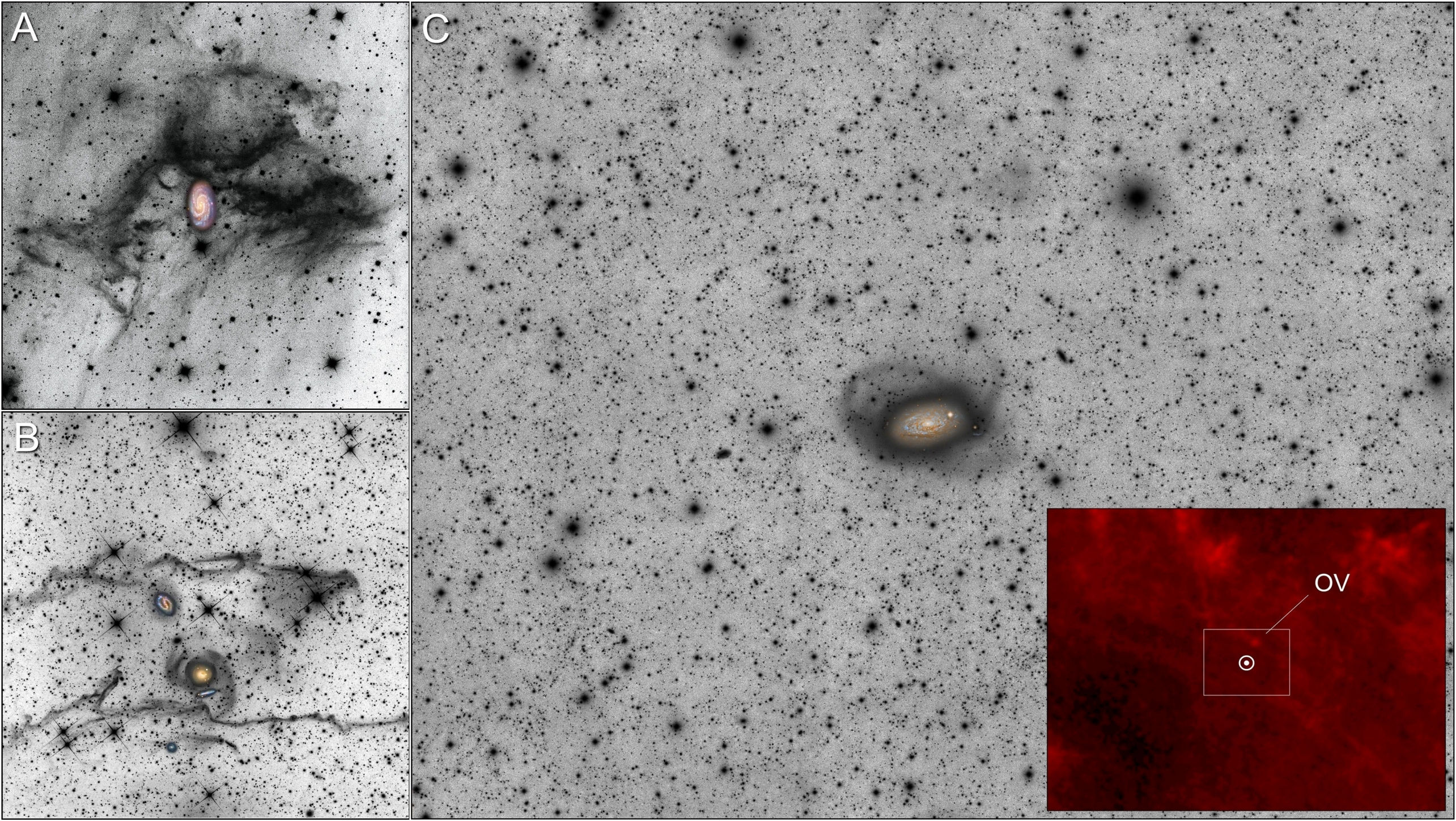}
    \caption{ ({\it left panels}): Examples of complex and filamentary Galactic cirri surrounding NGC 918 (panel A) and NGC 2634 (panel B), revealed in deep images obtained with Planewave RC 0.6-m  f/6.5 and Planewave 0.42-m f/6.8 telescopes, respectively. NGC 2634 also shows shell-like tidal features with similar shapes and widths, almost indistinguishable from the Galactic dust features. The FOV of these images are 20$\arcmin$ $\times$ 20 $\arcmin$ and 30$\arcmin$ $\times$ 30 $\arcmin$, respectively; ({\it right panel}) A deep image of the tidal stream around Messier 63, obtained with Takahashi FSQ106EDX f/5 telescope. A wide-field SFD dust map of this sky region (\citealt{SFD98, Schlafly2011}) is shown in the bottom right inset panel, with the corresponding FOV marked with a square. The comparison of our luminance filter image with this Galactic dust map shows that the low-surface brightness overdensity (marked OV) situated at 35$\arcmin$ NW of the galaxy is the brightest part of a cirrus filament, appearing as an isolated round over-density due to the surface brightness cutoff of our luminance image. }\label{fig:cirrus}
\end{figure*}


The presence of Galactic cirrus is probably one of
the most challenging issues in LSB research. While observations close to the Galactic disk have traditionally avoided high SB cirrus, for LSB observations the presence of cirrus at high Galactic latitudes is frequent \citep[e.g.,][]{2010MNRAS.403L..26C, 2010A&A...516A..83S, 2016ApJ...825...20B, 2018MNRAS.475L..40D}. The impact of the cirrus from the observational point of view is manifold. In Figure \ref{fig:cirrus} we present an example of cirri surrounding NGC 918, NGC 2634 and M63 (all not studied in this work) highlighting that cirrus can easily be confused with extragalactic features. Moreover, regions with considerable cirrus are susceptible to problems in sky background subtraction. This is because cirrus may fill a significant region of the study field (if not all), thus leaving few regions with no presence of sources for proper modeling and sky background subtraction. Currently, there are some proposed solutions to this problem, using as priors the IR Astronomical Satellite (IRAS; \citealt{IRAS}) mission or the Planck Space Observatory (\citealt{plank}) data to model the cirrus and sky background emission simultaneously \citep{2023ApJ...953....7L}, thus minimizing the impact on the data processing. However, this method is only effective in regions of high contamination by cirrus and on spatial scales similar to the resolution of the IRAS and Planck maps, on the order of 5 arcmin of FWHM \citep{2003NewAR..47.1017L, 2005ApJS..157..302M}. The other major problem is related to the confusion between cirrus and LSB structures or features. In this respect, there are some approaches followed recently. First, the modeling and subtraction of the cirrus emission using deep and high-resolution ESA Herschel Space Observatory \citep{2010A&A...518L...1P} data is possible (see \citealt{2017ApJ...834...16M}). However, this is not possible for the vast majority of fields due to the poor sky coverage of Herschel, and often lacks the necessary depth. Another approach is the use of optical colors to distinguish between extragalactic features and cirrus, according to the \textit{g}, \textit{r} and \textit{i} bands color characterization by \cite{2020A&A...644A..42R}. However, this requires the presence of at least 2 colors, in particular \textit{g-r} and \textit{r-i}, and is unfeasible for single-band photometric observations as is typically the case for the observations presented here.

In the absence of Herschel far-IR counterparts and multiple photometric bands, there are still possibilities to distinguish cirrus regions from genuine extragalactic features. The low-resolution IRAS and Planck maps are often useful to warn about the presence of cirrus clouds of large extent and relatively high density. For more filamentary features, visual identification is often a good resource. For example, clear connections of features to the central galaxy are highly likely to be streams. However, this classification is subjective, so there are no objective or analytical methods of discerning between cirrus and streams. For the case of dwarf galaxies, there are analytical approaches based on morphology that can be conclusive \citep{2025ApJ...979..175L}.

\section{Results}
\label{sec:Results}

In this Section, we analyse the deep images obtained on the galactic systems studied by the STSS (presented across Figures~\ref{fig:cmd1} to \ref{fig:cmd3}). For each galaxy, we focus our analysis on the presence of LSB structures and/or tidal perturbations in the galactic disk. 

\subsection{ESO~545-5}

\begin{figure*}
\centering
\includegraphics[width=0.7\textwidth]{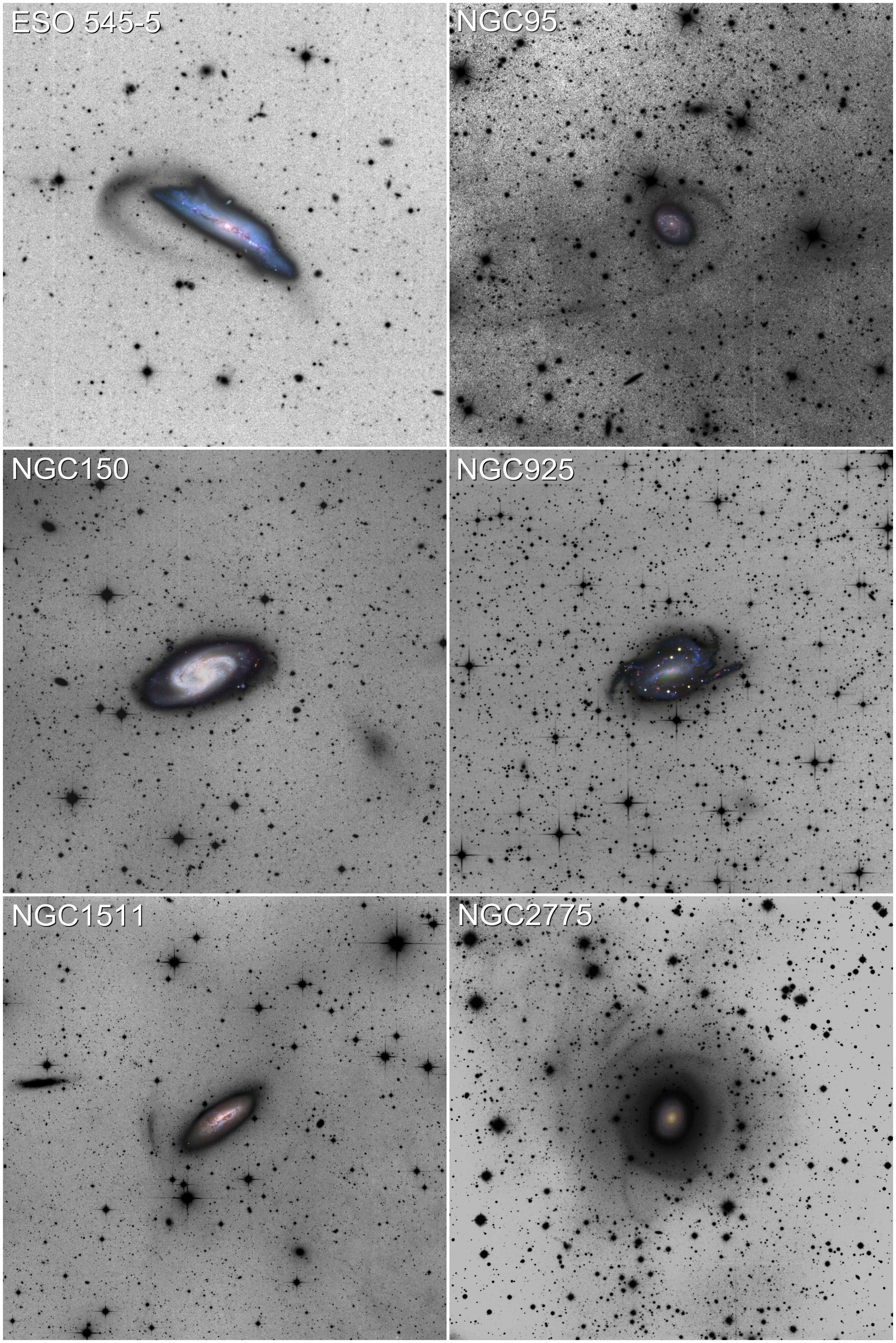}
\caption{Our STSS images of ESO 545-5, NGC 95, NGC 150, NGC 925, NGC 1511 and NGC 2775. North is up and East is left. See text for discussion on all six images.}\label{fig:cmd1}
\end{figure*}

Our image of ESO~545-5 (Figure~\ref{fig:cmd1}, top left) showcases a highly perturbed, edge-on disk, with a clearly discernible stream extending outwards from the northeast (NE) of the galaxy and looping around the southeast (SE). While difficult to see, it is plausible that the stream ends in the southwest (SW). From the image itself it is difficult to infer the overall morphology of the stream, although the observed fragment is consistent with a brighter loop of a rosette-like structure from a dwarf galaxy under disruption with a significant stellar mass (for example, >10$^{9}$ M$_{\sun}$). Indeed, a significant overdensity is observed in the loop, which could be the progenitor itself. Kinematic data are needed to better constrain the stream's morphology and properties, and thus the formation and perturbations suffered by this system. So far, the LSB outskirts of ESO~545-5 have not been studied in detail.



\subsection{NGC~95}

 NGC~95 (Figure~\ref{fig:cmd1}, top right) appears as a peculiar, barred spiral with a broken inner ring. Although the field is significantly contaminated by Galactic cirrus, our image showcases a long, loop-like tidal structure that seems to extend out of our field of view, potentially formed by the infall of a dwarf satellite galaxy. Towards the west of the galaxy, the stream appears to show a discontinuity; interestingly, the propagation of the stream through the discontinuity does not seem to connect both fragments. It is therefore difficult to assert if the full feature has a common origin from a single satellite, or if there are two different pieces of streams overlapping on the sky projection. In the case of the former, it is not well understood how an interaction between the host galaxy and an infalling satellite could induce an abrupt change of direction of the satellite. While an additional gravity source could have created a gap in the stream (see e.g., \citealt{Erkal2015a}), additional data are required to discern the origin of the discontinuity. See \cite{MD2023SSLS} for a follow-up SSLS image of NGC 95.

\subsection{NGC~150}

\begin{figure*}
  \centering
    \includegraphics[width=0.80\textwidth]{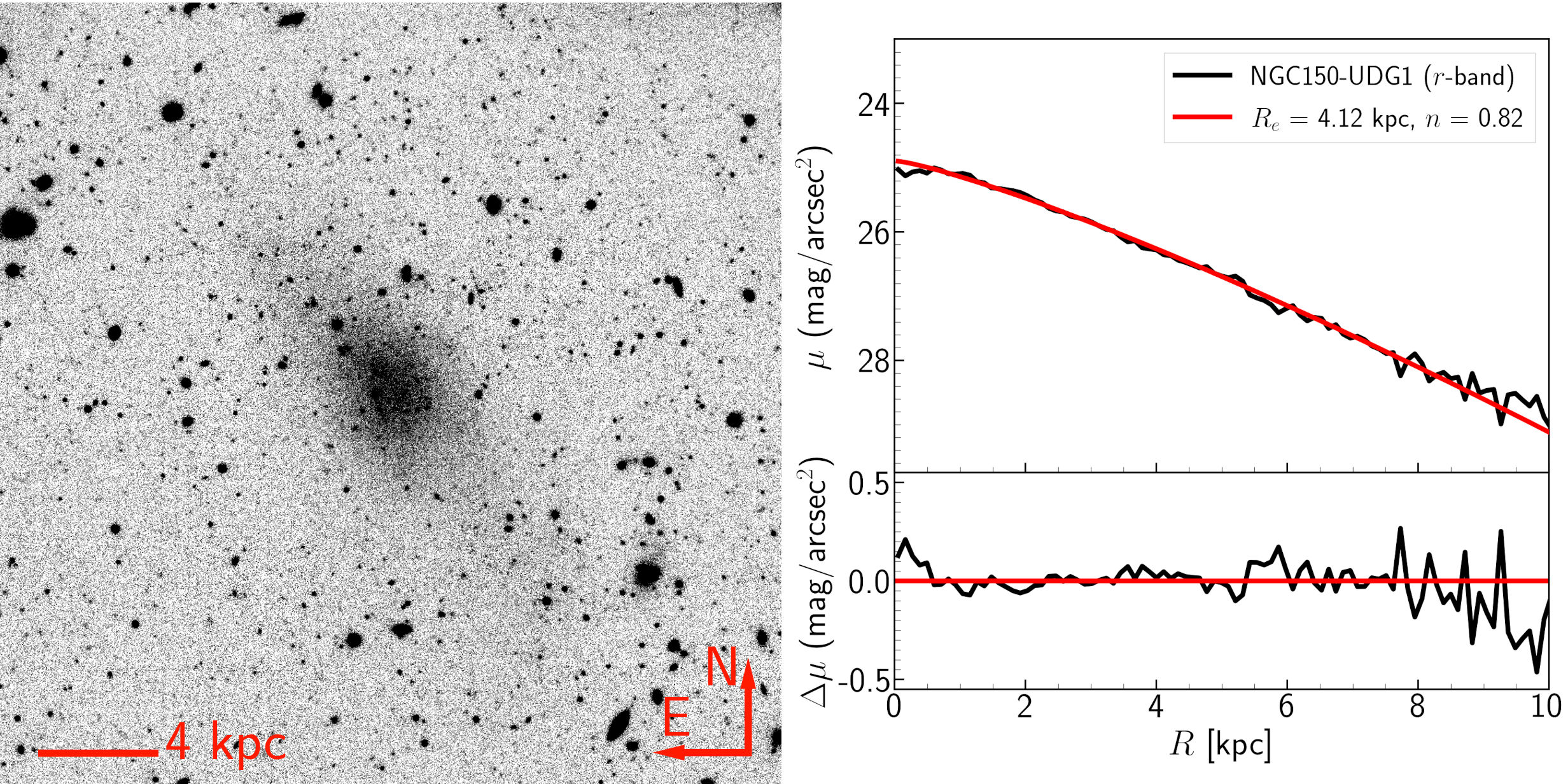}
    \caption{\textit{Left panel}: the galaxy NGC150-UDG1 in the $r$-band. \textit{Right panel}: the light profile (black line) and the best-fit S\'ersic profile (red curve) for the galaxy. The corresponding S\'ersic parameters of the best-fit function are presented on top right of the diagram.
    }\label{fig:ngc150-udg1-r}
\end{figure*}

\begin{table*}
\centering
\caption{The structural parameters of NGC150-UDG1. Columns from left to right represent (a) the observed filter, (b) effective radius, (c) S\'ersic index, (d) major axis position angle measured counterclockwise from north to east, (e) ellipticity, (f) total apparent magnitude, (g) total absolute magnitude, and (h) surface brightness within one effective radius (i.e. the mean effective surface brightness).}
\begin{tabular}{ lccccccc } \hline  
Filter & $R_{\rm e}$ & $n$ & PA & $\epsilon$ & $m$  & $M$  & $<$$\mu_{\rm e}$$>$ \\
 - &  kpc & - & deg & - & mag & mag & mag/arcsec$^2$ \\
(a) & (b) & (c) & (d) & (e) & (f) & (g) & (h) \\
\hline
$g$ & 3.90 & 0.86 & 31.85 & 0.58 & 16.78 & -14.87 & 26.64 \\
$r$ & 4.12 & 0.82 & 33.160 & 0.55 & 16.38 & -15.27 & 26.35 \\ 
\hline 
\end{tabular}
\end{table*}\label{sersicparams}

\begin{figure*}[hbt!]
\centering
	\includegraphics[width=0.9\textwidth]{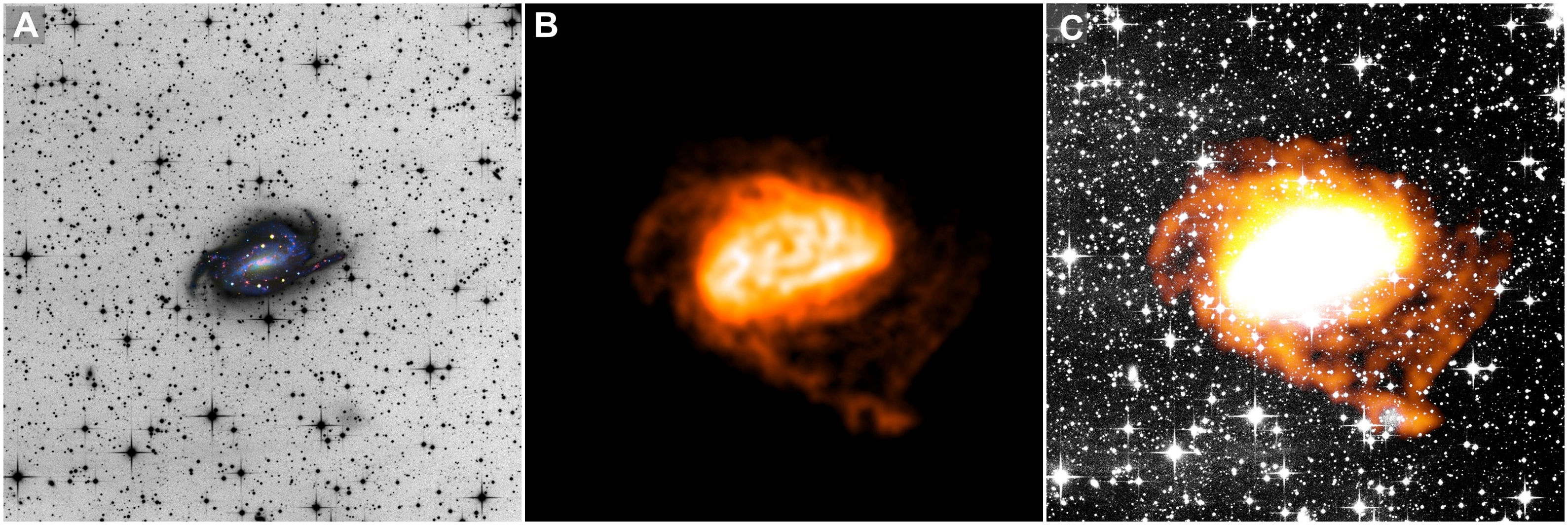}
    \caption{A comparison of our NGC~925 image (panel A) against a H$\textsc{i}$ map taken from the HALOGAS survey (panel B). The low resolution datasets (with a synthesized beam of 37$\,.\!\!^{\prime\prime}$9 $\times$ 33$\,.\!\!^{\prime\prime}$2) were used to recover the diffuse gas. In panel C we overlay the H$\textsc{i}$ map on our image. 
}\label{fig:NGC925}
\end{figure*}

Our deep images of NGC~150 (Figure~\ref{fig:cmd1}, middle left), previously unreported in literature, reveal a large, asymmetric LSB overdensity to the west of the galaxy. If the overdensity is at the same distance as NGC~150, then its size and SB (given in Table \ref{sersicparams}) could be an ultra diffuse galaxy (UDG). To further examine this LSB galaxy candidate, we analysed its light profile and obtained its S\'ersic parameters (shown in Fig. \ref{fig:ngc150-udg1-r}) using the DES data release 2 (DR2) data in the \textit{g} and \textit{r} band. After an initial background subtraction of the tiles, we cut out a region of 1200$\times$1200 pixels centred on the galaxy in both bands, corresponding to the physical size of $\sim$30$\times$30 kpc at the distance of the galaxy (on the assumption it is associated with NGC 150 at 20.9 Mpc). We used the cropped frames as the main science frame for the analysis using \textsc{GALFIT} (\citealt{Peng2002}), additionally preparing a sigma frame, a bad pixel map (mask) and a point-spread function (PSF) for the input. Figure \ref{fig:ngc150-udg1-r} shows the object in the r band (left), and the galaxy light profile and the best-fit model is shown (right). The final S\'ersic parameters are presented in Table 5.

Given the effective radius and SB of this object, at 20.9 Mpc (3.90 kpc and 26.64 mag/arcsec$^{2}$ in g-band), the galaxy satisfy the UDG criteria (\citealt{vanDokkum2015}). The large effective radius of the galaxy puts it among the most diffuse UDGs known. The elongation of this UDG candidate (hereafter NGC150-UDG1) towards NGC 150 suggests an ongoing tidal interaction which could be connected to the formation of this UDG. Such tidal features has been seen in several other cases in similar environments (\citealt{Bennet2018, zemaitis2023}) and are suggested to be responsible for transformation of non-UDG dwarf galaxies to UDGs, in particular in higher density environments (\citealt{Sales2020}). This formation channel, as discussed in \cite{Carleton2021} also suggests that UDGs formed through tidal interactions would host more globular clusters (GCs) than non-UDG dwarf galaxies of a similar stellar mass. The small and point-like sources in the central regions of NGC150-UDG1 could be GCs, however given the quality of the data and the distance of this galaxy, an in depth analysis of the GCs is not possible. Given the current coverage map of the Euclid mission (\citealt{Mellier2024}), the galaxy will be observed by Euclid which allows detailed analysis of the tidal features in optical and near-infrared (\citealt{Urbano2024}) as well as the GCs (\citealt{Saifollahi2024}) at this distance.

\subsection{NGC~925}

\begin{figure*}
    \centering
	\includegraphics[width=0.7\textwidth]{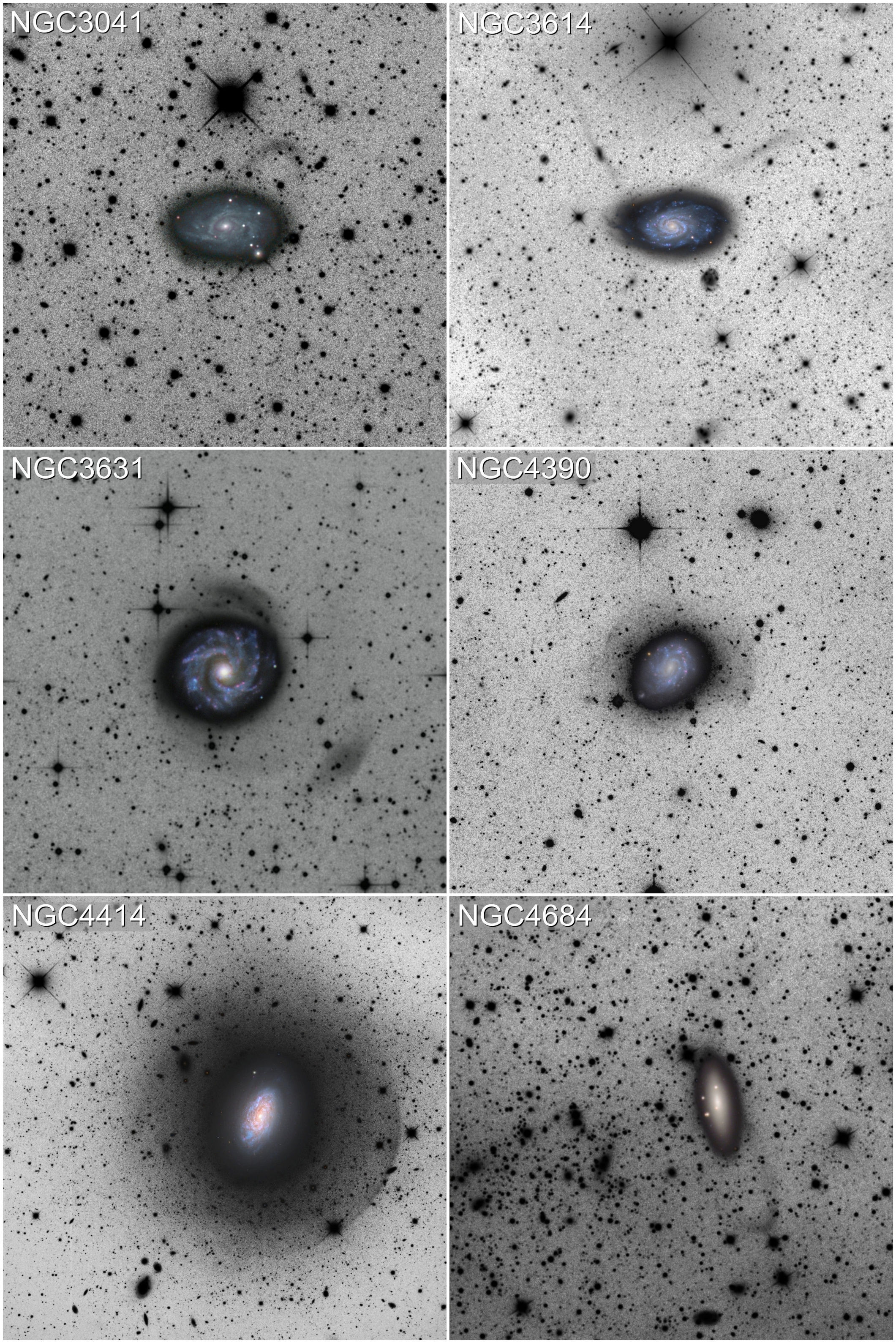}
    \caption{Our STSS images of NGC 3041, NGC 3614, NGC3631, NGC 3631, NGC 4390, NGC 4414 and NGC 4684. See text for discussion on all six images.}\label{fig:cmd2}
\end{figure*}


We report the first deep image of NGC 925 (Figure~\ref{fig:cmd1}, middle right). This galaxy showcases two prominent spiral arms in a backwards S shape, with no galactic bulge at the centre but a faint, yet well-defined, bar. The SE arm extends further than its northwestern (NW) counterpart, with the tip hosting prominent stellar associations, while the NW arm appears somewhat diffuse and disturbed. The bar, which is offset from NGC~925's center by 1 kpc (\citealt{Pisano1998}) contains many H~II regions and resolved stars (visible due to its proximity of $\sim$ 9 Mpc), and has been previously found to host widespread star formation even in areas with low gas density \citep{Pisano2000}. Our image confirms an extended disk, with long star-forming spiral arms that reach the disk’s outskirts. We can clearly see how the main spiral arms are connected to the central stellar bar, suggesting they could have been induced by bar's invariant manifolds \citep{RomeroGomez07}. 
Interestingly, in the SW of the galaxy we observe a previously unreported LSB clump which, due to its unresolvable faint nature, we cannot confirm if it is a neighbouring LSB galaxy with our photometry alone. However, in Figure \ref{fig:NGC925}, we traced this structure using HI velocity-integrated emission (zeroth moment) data from the HALOGAS survey and demonstrated a link between the overdensity and NGC~925 in the form of a gas bridge.



\subsection{NGC~1511}



The starbursting galaxy NGC~1511 (Figure~\ref{fig:cmd1}, bottom left) is the largest member of a compact group of interacting galaxies. In spite of its edge-on orientation, \citet{Buta2007} suggest it could be a late-type spiral galaxy with a single spiral arm, and a bright nucleus \citealt{Querejeta2021}). Our data shows the spiral arm (SE of the galaxy) to host bright star-forming knots. The star-forming disk appears dusty with extraplanar gas, and the SW side of the bulge is obscured by a dust band, making the central region appear indented. Our deep optical image confirms a faint arc SE of the galaxy, which we tentatively link to its interaction history with its neighboring galaxies (NGC 1511A and NGC 1511b) in agreement with studies of NGC~1511 in X-ray, H$\alpha$, UV and near-infrared (e.g. \citealt{Dahlem2003}). While not shown in their figure, the authors comment on the presence of a tidal tail (SE to north) in their X-ray data, concluding that the galaxy is heavily disturbed on its eastern side. We confirm this result with our complementary optical image.




\subsection{NGC~2775}

\begin{figure*}
  \centering
    \includegraphics[width=0.80\textwidth]{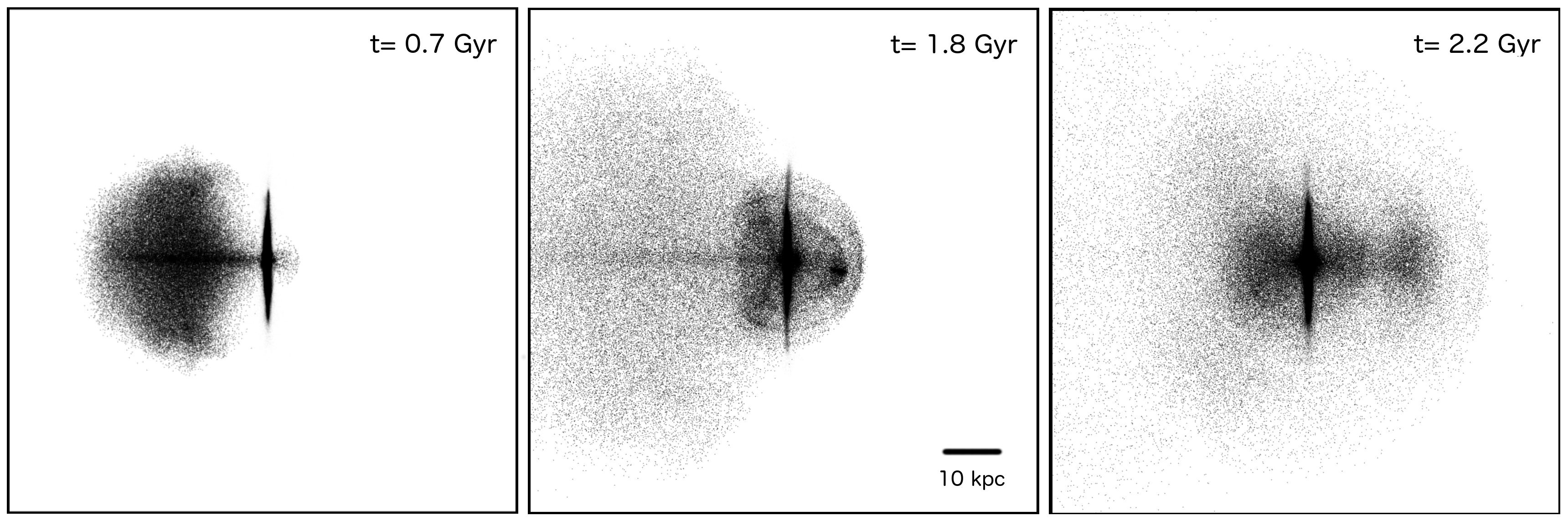}
    \caption{N-body model of a minor merger between an NGC~4414-like galaxy and a low-mass dwarf galaxy in an on-center interaction. From left to right we show the evolution of the system during 2.2 Gyr. The model has been obtained following a process similar to the one in \citet{NGC922-2023}}.
    \label{fig:NbodyNGC4414}
\end{figure*}

\begin{figure*}
\centering
	\includegraphics[width=0.7\textwidth]{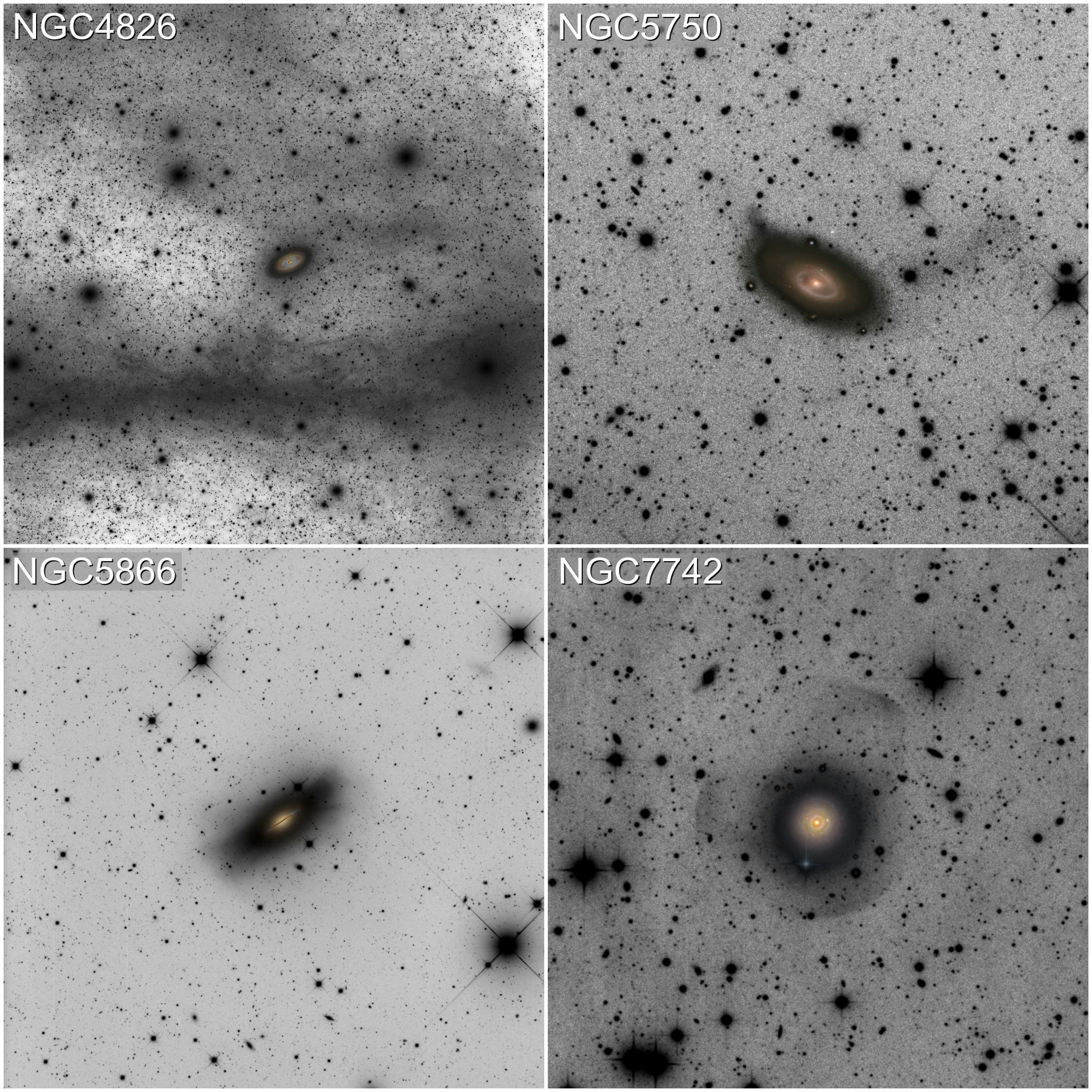}
    \caption{Our STSS images of NGC4826, NGC5750, NGC5866 and NGC7742. See text for discussion on all four images.}\label{fig:cmd3}
\end{figure*}

\begin{figure}
\centering
	\includegraphics[width=0.9\columnwidth]{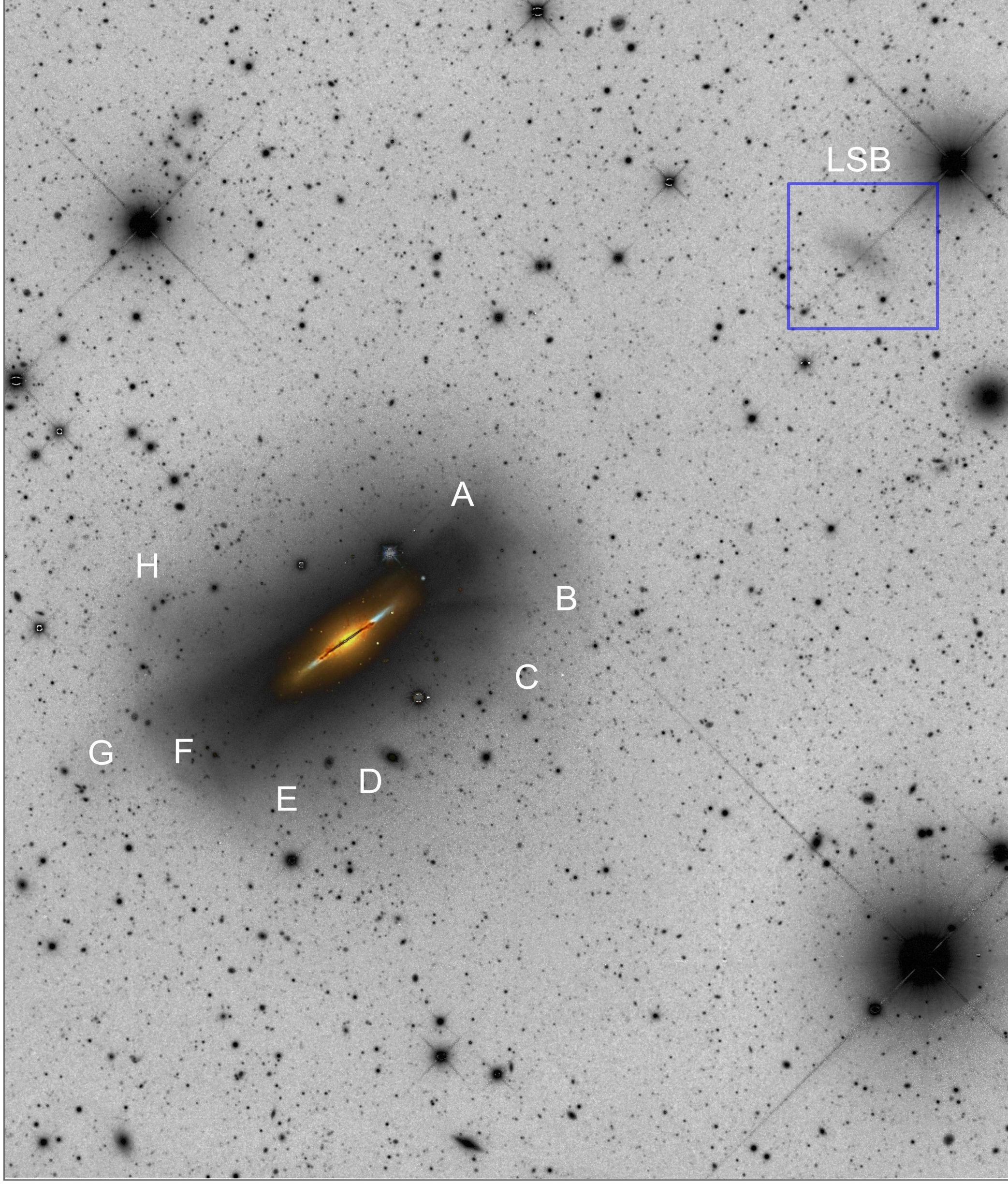}
    \caption{ Stellar features detected in the NGC~5866 inner halo. They include a conspicuous wedge in the NW side of the disk that suggests a recent dwarf accretion (A), coherent "spikes" (B,E, F, G), a very faint round over-density (C) and fuzzy outer plumes (D,H). The FOV is 20$\arcmin$ $\times$ 25 $\arcmin$. }
    \label{fig:NGC5866}
\end{figure}

NGC 2775 (Figure~\ref{fig:cmd1}, bottom right) is a member of a small group of galaxies (alongside NGC 2777, NGC 2773) and part of the Virgo Supercluster. NGC 2775 showcases a smooth, elliptical-like bulge surrounded by a tightly wound, segmented spiral structure that extends around the disk, with a dust ring encircling the entire galaxy. Our deep image showcases three, large scale shell structures at different radii on the NE/N side of the galaxy, for the first time. The symmetrical nature of the features suggests that a past satellite merger endured more than one apocentre on its orbit around its host, although it remains possible that the innermost ring could be an independent feature. In the SE, we observe a loop feature extending out of the galaxy, which could also be related to a different accretion event.

\subsection{NGC~3041}

NGC 3041 (Figure~\ref{fig:cmd2}, top left), a relatively isolated galaxy hosting an AGN, showcases a very compact galactic core with a spiral structure that extends out of the central region. Further out, the galaxy displays numerous spiral arms hosting prominent star-forming regions. The western arms, however, appear more symmetrical than those on the eastern side, one of which appears to jut almost straight out from the central coil. In our image, we observe a thick and diffuse loop-like stream NW of the galaxy, indicative of the apocentre of a past satellite merger event.


\subsection{NGC~3614}

NGC 3614 (Figure~\ref{fig:cmd2}, top right) is a barred spiral galaxy, with two main arms branching outwards from an internal ring, and thin, spiral arms containing numerous star forming regions. We note that the small spiral galaxy SW of NGC 3614 is a background galaxy. Our deep image identifies, for the first time, two tidal streams which are completely disconnected and probably correspond to two different accretion events. One of the streams (NE) shows the tidal disruption of its progenitor, with both the leading and the trailing tidal tails.

\subsection{NGC~3631}


NGC 3631 (Figure~\ref{fig:cmd2}, middle left) is an isolated spiral galaxy, with a small, bright nucleus, extended stellar disk, tightly coiled spiral arms, and a symmetrical appearance overall. With the first deep image of NGC 3631, we report a diffuse LSB `arm' north of the galaxy and a significantly fainter, diffuse arm-looking feature south of the galaxy. We additionally note a shell-like overdensity of debris SW of the galaxy.





\subsection{NGC~4390}

NGC 4390 (Figure~\ref{fig:cmd2}, middle right), located in the Virgo Cluster, is a barred spiral galaxy, with star forming spiral arms extending all the way from the nucleus to the outskirts. SE of the galaxy (shown within the colour image) lies a candidate satellite dwarf galaxy. Our image highlights beautiful shells surrounding the galaxy, each located at different radii from centre- from west to east we can faintly trace two shells, indicative of a past merger. See \cite{MD2023SSLS} for a follow-up SSLS image of NGC 4390.



\subsection{NGC~4414}
NGC 4414 (Figure~\ref{fig:cmd2}, bottom left) is an unbarred, isolated spiral galaxy in the Coma I Cluster showcasing blue spiral arms blooming with ongoing star formation. Our data clearly map an extended stellar halo surrounding the galaxy, and a well-defined shell-like structure in the SW (additionally traced by HALOGAS; \citealt{2014A&A...566A..80D}). Such halo structures can signal a recent minor merger with a low impact parameter (see, e.g. \citealt{NGC922-2023}). To reinforce this scenario, we created an N-body model of NGC 4414 recently experienced a minor merger. We used the same properties as those adopted by the model presented in \cite{NGC922-2023}, however, now with an impact parameter equal to zero to mimic the formation of the structures observed in NGC 4414 after experiencing a minor merger. For this model, we used a mass resolution of 2.5$\times$ 10$^4$ M$_{\sun}$ for both the dark matter and stellar particles, and a spatial resolution (minimum AMR cell size) of 40 pc. The simulation volume is a box with 1 Mpc/side with h=0.7. The central galaxy is simulated as a stellar exponential disk embedded in a Navarro-Frenk-White (NFW; \citealt{NFW1997}) dark matter halo. The initial conditions of the collisionless components were obtained using the Jeans equation moments method as introduced by \citealt{Hernquist1993}. The compact dwarf system was simulated as a extended distribution of particles, which allowed us to reproduce the observed stellar stream. The initial condition of this system is a simple stellar structure that follows a compact NFW profile\footnote{See Table 2 in \cite{NGC922-2023} for the parameters used to generate both the central galaxy and the compact dwarf in this simulation.}. 

In Figure~\ref{fig:NbodyNGC4414} we show three snapshots of the simulation. In the first panel to the left we show the first shell created after the first apocenter, in the central panel we observe the formation of concentric shell-like structure (very similar to those observed in NGC 4414), including the differences in symmetry between the infalling satellite’s direction of approach and the splash-back observed in Figure\ref{fig:cmd2}; finally, in the right panel, we show the almost phase mixed stellar halo structure resulting from the disruption of the dwarf. With these three panels we show the possible evolution of the system over a period of 2.2 Gyr.

\subsection{NGC~4684}

NGC~4684 (Figure~\ref{fig:cmd2}, bottom right) appears as an elongated elliptical galaxy, with a diffuse nucleus encompassed by a dust ring, (which, in turn, is surrounded by a bright ring). The galaxy appears to be experiencing relatively low star forming activity given its lack of bright star formation regions. Our images confirm a large, loop-like feature in the SW which appears to be coplanar. This feature has already been noted by \citealt{Miskolczi2011} (SDSS) and \cite{Mancillas2019} (MATLAS survey; \citealt{MATLAS}).

\subsection{NGC~4826}

\begin{figure}
  \centering
    \includegraphics[width=0.90\columnwidth]{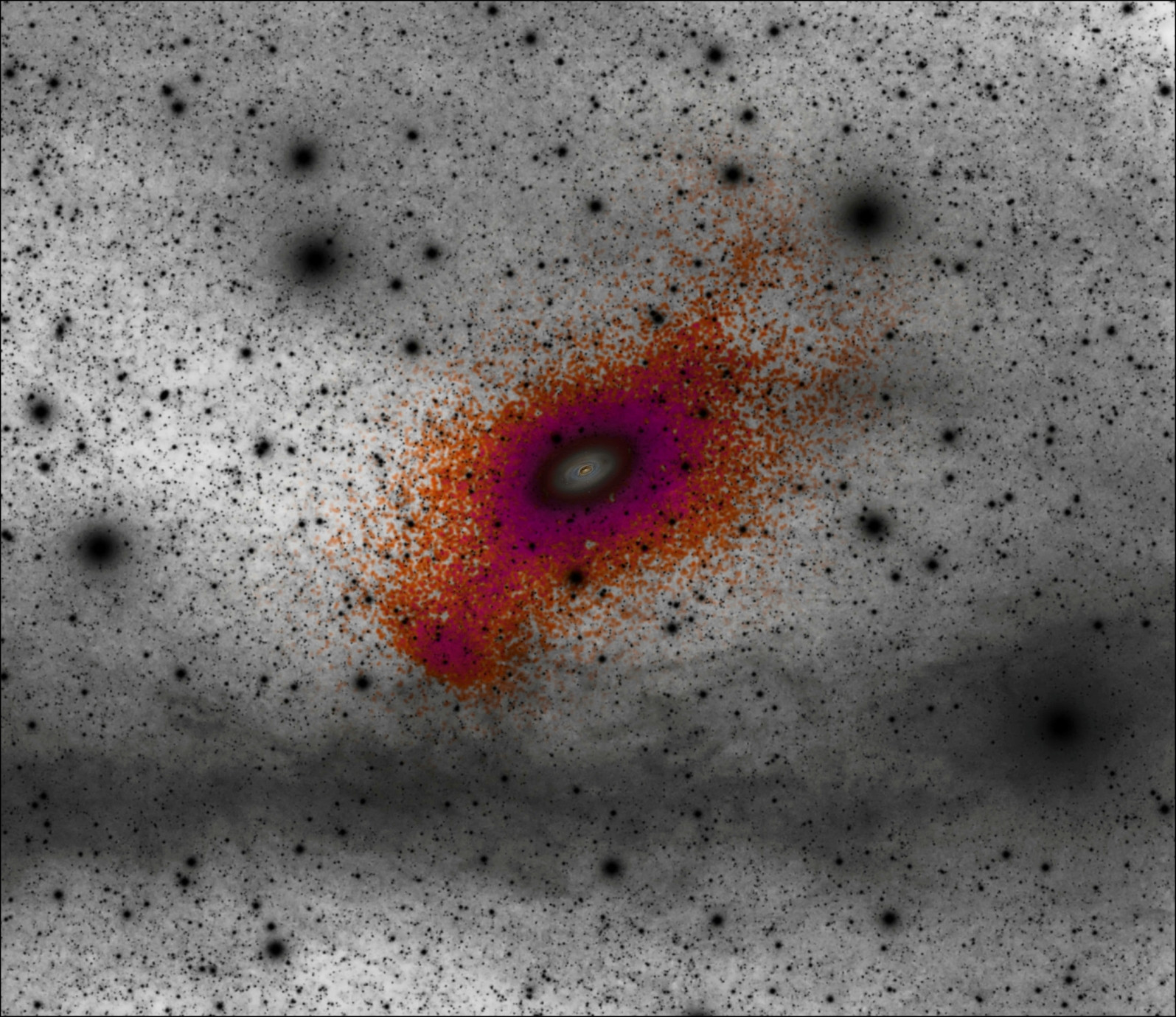}
    \caption{A comparison of our deep image of NGC~4826 with the
    stellar density map from \cite{Smercina2023}. We can clearly discern that the LSB features, which blend with galactic cirrus in the original image, are stellar debris.}
    \label{fig:NGC4826}
\end{figure}

NGC~4826 (Figure~\ref{fig:cmd3}, top left, also known as M64) exhibits strong dust extinction across its central regions which, thanks to the contrast achieved against its virtually dust-free outer disk, earned the galaxy nicknames such as the `Black Eye', `Evil Eye' or even `Sleeping Beauty'. The galaxy's high dust extinction in the central regions has been liked to a past merger with a massive, gas-poor disk galaxy (\citealt{Braun1992, Braun1994, Rix1995}). At first glance, our deep image of NGC~4826 appears highly contaminated with galactic cirrus and the LSB structures do not seem credible. However, in Figure \ref{fig:NGC4826}, we compare our deep image of NGC~4826 with the stellar density map from \cite{Smercina2023} demonstrating an abundance of stream features and, most notably, we recover an impressive umbrella-like feature in the SE. In the NW we see two (potentially three) stream features which, while they could also be umbrella streams, show a  morphology that is harder to classify.



\subsection{NGC~5750}

NGC 5750 (Figure~\ref{fig:cmd3}, top right) is a barred spiral galaxy with a subtle, inner ring, showcasing dusty filaments branching north of the nucleus. These features are set within a blue ring of (seemingly overlapping) star forming regions and, beyond it, the disk becomes increasingly laminar. Our image reveals a few points of interest- first, we observe a truncated overdensity NE of the disk. NW of the galaxy, we detect a 'smoke-like' diffuse feature, which could be residue of a past accretion event. It is unclear whether these features could be caused by the same accretion event. These features have been noted in shallower detail in enhanced SDSS imaging (see \citealt{Morales2018}). A SB measurement of the stream from DESI LS data is available in \cite{Miro-Carretero2023}.



\subsection{NGC~5866}

NGC 5866 (Figure~\ref{fig:cmd3}, bottom left, also known as M102) exhibits a narrow dust lane that is often in the glare of the galaxy’s bulge and wing-like halo, hence its nickname as the Spindle Galaxy. Beyond the dust lane, at opposite ends, the galaxy’s outer disk is defined by very sharp beams projecting outward. At first glance, the disk appears thick with no signs of recent interactions with satellites. However, as we show in Figure \ref{fig:NGC5866}, resolving the halo reveals complex substructures, which may have been caused by old accretion event(s). We also annotate a LSB feature approximately 10 arcmin from the center of the galaxy, identified as KK 236 in \cite{Karachentseva1998}. It is unlikely that the feature could be linked to an old accretion event, given that the distance between NGC 5866 and KK 236 is over three times the size of the halo features. Additionally, \cite{Huchtmeier2000} measured HI emission and a negative radial velocity for KK 236 ($V_{rad}=-150$ km/s), suggesting that this feature is just a local HI cloud. 

STSS images of NGC 5866 have been previously presented in \citealt{MD2010}. Deep luminance filter images of NGC 5866's halo have also been obtained by the HERON project (\citealt{Mosenkov2020}), concluding an `oval/boxy' shape, with hints of filamentary structures within the halo (which we resolve in Figure \ref{fig:NGC5866}). We note \citealt{Lanzetta2024} also obtained luminance filter images of the NGC 5866 group (not the galaxy itself, see their fig. 7). NGC 5866 has also been imaged by \cite{MD2023SSLS} (DESI LS).

\subsection{NGC~7742}

NGC~7742 (Figure~\ref{fig:cmd3}, bottom right) showcases two blue rings, indicative of active star formation. Interestingly, over three decades ago, the formation of galaxies with inner H II rings was linked to past encounters where the galaxy experienced an almost direct collision with the incoming galaxy (such that it penetrated the centre of the other). In our images, we observe several tidal shells surrounding the galaxy. These results are compatible with a scenario where a satellite interacted with the central galaxy (with a small impact parameter), perturbing the disk and triggering the formation of consecutive tidal rings within the galaxy. In the CDM universe where the rate of minor mergers is high, the theory predicts that the majority of disk galaxies should show blue star forming rings (collisional rings) over a Hubble time (see \citealt{Theys1977}). These structures are produced by the impact of a dwarf galaxy with the disk of the central galaxy, following an almost radial orbit. The impact generates a compression and a subsequent expansion of the “stellar fluid” that propagates across the disc as a tidal ring (\citealt{Lynds1976, Hernquist1993}). If the disk of the central galaxy is gravitationally unstable (e.g., has a large gas fraction), becomes also a propagating star forming ring (blue ring, see e.g. Cartwheel galaxy, \citealt{Higdon1996}). The lifetime of these structures has been studied using simulations and semianalytical models it was concluded that these star forming rings only lasts for $\sim$ 0.2-- 0.5 Gyr (\citealt{Wong2006, Pellerin2010, Renaud2018, Elagali2018}). After this period they become only marginally observable up to 0.7 Gyr after the collision (\citealt{Wu2015}). Due to the short lifetime of these structures, the frequency of observations in nearby galaxies with stellar streams can provide us with a new tool to constrain the nature of dark matter when combined with the prediction from semi-empirical models of the number of expected minor in different dark matter paradigms. However, it is important to ensure that the star forming rings produced by the rejuvenation of the central galaxy must be excluded from this statistics (e.g., \citealt{Silchenk2023}). Future large surveys of the LSB universe like the STSS (combined with deep ultraviolet data) will open the door to these new tests on the dark matter nature. NGC 7742 has also imaged by \cite{Morales2018} (SDSS) and \cite{Rich2019} (HERON).

\subsection{NGC~2460}

\begin{figure}
	\includegraphics[width=0.99\columnwidth]{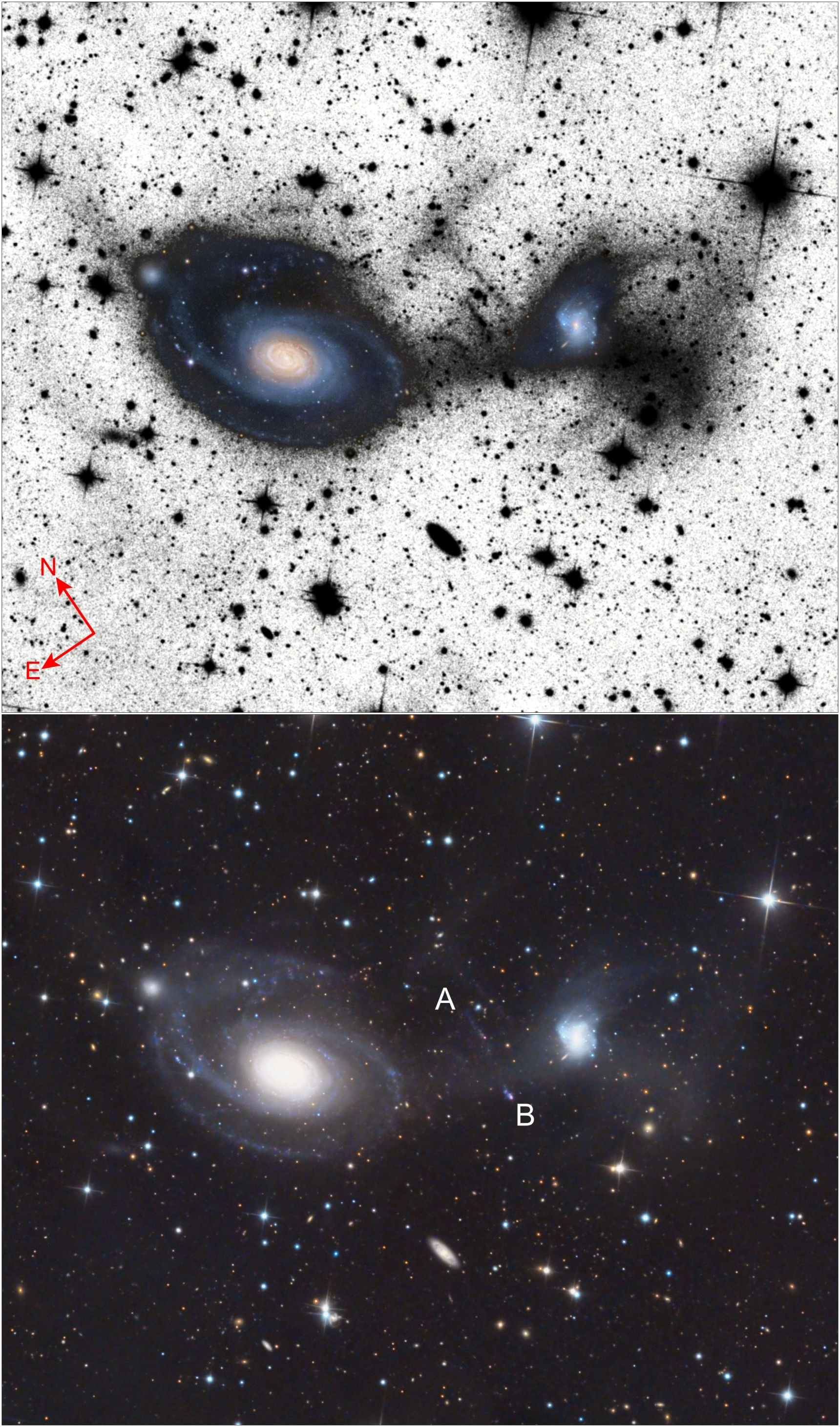}
    \caption{ \textit{Top panel:} Luminance-filter, linear-stretched image of NGC~2460 and IC~2209 and their tidal features obtained with the BBO~II RCO 0.506-m f/8.1 telescope \textit{Bottom panel:} Full color image obtained with Mount Lemmon Sky Center 0.80-m f/7 telescope. The two emission over-densities in the intersection between the tidal bridge and the NGC~2460 disk are marked with A and B.}
    \label{fig:NGC2460}
\end{figure}

Finally, we showcase our deep imaging results for NGC~2460 (Figure~\ref{fig:NGC2460}). Classified as an AGN (\citealt{2011A&A...534A..31G}), NGC~2460 displays a bright, concentrated core and a dusty, tightly wound spiral structure extending to the nucleus. The outer disk forms two arms that appear to merge into one (see also \citealt{2011AJ....142..145G}).  Seemingly nearby galaxies include the unrelated LEDA 213434 and the interacting IC~2209, whose similar redshift distances ($\sim$30\,Mpc) suggest their interaction may have created NGC~2460's extended tidal arms \citep{2024MNRAS.527.9605C}. The outer disk forms two arms that appear to merge into one \citep{2011AJ....142..145G}. 
NGC~2460 is part of a dense group with two main interacting disk galaxies, a spheroidal dwarf, and various satellites with tidal features. In addition to a clear bridge, the smaller disk galaxy shows significant disruption, with tidal structures likely from smaller satellite destruction, including a "dog leg" (similar to that observed in the NGC 1097 stellar halo; \citealt{Amorisco2015}) or umbrella and a tidal structure perpendicular to the bridge. GALEX images reveal two star-forming clumps within this bridge (which we mark as A and B in the Figure), but without kinematic data it is difficult to conclude if they are star formation regions related to the outer arm of NGC~2460, or evidence of star formation enhancement in the bridge due to the tidal interaction between both galaxies (e.g. see \citealt{Beaton2014}). A follow-up spectroscopic study of these two star formation knots in the tidal bridge's structure is planned as a future work. NGC 2460 has also been images by the DESI LS in \cite{MD2023SSLS} and a SB measurement of its stream is also available in \cite{Miro-Carretero2023}.

\section{Conclusions and future work}
\label{sec:Conclusions}

The STSS has yielded an unprecedented sample of bright stellar streams in nearby spiral galaxies, including the discovery of observational analogues to the canonical morphologies found in cosmological simulations of stellar halos. Amateur astrophotography has demonstrated its robust, impressive capabilities in studying galaxy assembly in the local Universe, permitting scientific analysis of LSB galactic features with low-cost equipment.

In this paper, we have provided deep imaging of 15 nearby massive galaxies achieved with amateur telescopes. We have revealed new, previously unseen features related to the recent dwarf satellite accretion in their halos, including other phenomena, previously overlooked in shallower large-scale photometric surveys. Our main conclusions are as follows:

  \begin{enumerate}
     \item Modern amateur telescopes are capable of producing excellent deep, wide-field imaging of nearby galaxies, with depths comparable to professional projects. Even without an optimized data acquisition strategy (e.g., accurate flat field calibration, dithering pattern, etc., employed by professional telescopes), modern amateur telescopes can obtain images reaching SB limits of $\sim28$ mag/arcsec$^{2}$, comparable to more expensive professional projects (such as DESI LS, Dragonfly, see Fig. \ref{fig:sblimit_hist}), although deeper results are possible depending on the exposure time adopted and other factors. We note that the addition of a data acquisition strategy makes it feasible to reach even deeper (30 - 30.5 mag/arcsec$^{2}$) with this low-cost equipment (\citealt{2023A&A...679A.157R}).
     
    \item Our amateur data illustrate how deep imaging of galactic outskirts can shed new insights on a wide range of aspects on their formation and evolution. We have discussed features such as: (i) the (potential) star formation enhancement in galactic disks due to tidal stream interaction (e.g., NGC 2460; NGC 7742), (ii) giant, puffy satellites with tidal tails (e.g, NGC 150), (iii) an extended gas bridge linked to a stellar over-density (e.g., NGC 925), (iv) provided insights into ISM outflows (e.g, NGC 5750), and (v), discussed stellar stream activity.

   \item Robotic amateur telescopes are excellent facilities for wide-field imaging (>1 deg) around nearby galaxies (< 25 Mpc). Along with their very low cost, they avoid the issues associated with 'mosaicing' smaller images to create a large, wide-field image (e.g. see Fig. \ref{fig:mosaic}) which may occur given that, among other factors, each frame has been taken under slightly different sky conditions.  Due to these inconsistencies, the final image suffers unintentional removal of significant faint features during the background subtraction process. 

  \end{enumerate}

During the upcoming years, the STSS will focus on identifying nearby (< 25-30 Mpc) stellar streams that, based on their properties (surface brightness, morphology, orientation, etc.), represent the most promising targets for stellar population and dynamical studies with the new generation of large telescopes and space instruments becoming available in the next decade (e.g., LSST, Roman, ARRAKIHS).\\ 


Looking forward, amateur astrophotography is transitioning from using CCDs to CMOS sensor technology. While CCD technology has produced reliable results with good sensitivity, it has been rendered obsolete by several factors: its expensive manufacturing costs in comparison to CMOS, its large pixel size (>6$\mu m$) becoming increasingly incompatible with the short focal length telescopes presently favoured, and its inability to compete with the continuously improving specifications of CMOS alternative. As a result, astronomical camera manufacturers have stopped producing CCD-based models. Modern CMOS sensors offer substantially reduced read and thermal noise (typically 2-5 times lower than CCDs), smaller sensor sizes which greatly improve resolution with short focal lengths, and back-side illuminated (BSI) designs that enhance quantum efficiency by positioning the circuit wiring behind the photosites, creating an unobstructed path for incoming photons and improving performance in low-light conditions.

While early CMOS sensors with 12-bit and 14-bit analog-to-digital converters (ADCs) raised legitimate concerns about their suitability for scientific applications due to their limited dynamic range (representing only 4,096 and 16,384 gray levels respectively), the latest generation of sensors (with a 16-bit ADC) match the 65,536 gray level capability of scientific-grade CCDs. These advancements have largely addressed the linearity concerns that initially made astronomers hesitant to adopt CMOS technology for scientific work. The exceptional linearity demonstrated by modern sensors (e.g., Sony’s IMX sensors, see \citealt{Alarcon2023}) suggests that high-end CMOS cameras satisfy the rigorous requirements for quantitative astronomical measurements that were once the exclusive domain of CCD instruments, simultaneously offering superior noise characteristics, higher quantum efficiency, and significantly lower cost. In this context, future releases of STSS will utilise CMOS-based telescopes and explore their suitability for LSB imaging of galaxies.



\begin{acknowledgements}
DMD acknowledges the grant CNS2022-136017 funding by MICIU/AEI /10.13039/501100011033 and the European Union Next Generation EU/PRTR, the financial support from the Severo Ochoa Grant CEX2021-001131-S funded by MCIN/AEI/10.13039/501100011033 and project PDI2020-114581GB-C21/ AEI / 10.13039/501100011033. DMD acknowledges the financial support provided by the Governments of Spain and Arag\'on through their general budgets and  Fondo de Inversiones de Teruel, and the Aragonese Government through the Research Group E16\_23R.

DJB acknowledges funding from the German Science Foundation DFG, within the Collaborative Research Center SFB1491 "Cosmic Interacting Matters - From Source to Signal". MAGF acknowledges ﬁnancial support from the Spanish Ministry of Science and Innovation through the project PID2022-138896NB-C55

 The Pan-STARRS1 Surveys (PS1) and the PS1 public science archive have been made possible through contributions by the Institute for Astronomy, the University of Hawaii, the Pan-STARRS Project Office, the Max-Planck Society and its participating institutes, the Max Planck Institute for Astronomy, Heidelberg and the Max Planck Institute for Extraterrestrial Physics, Garching, The Johns Hopkins University, Durham University, the University of Edinburgh, the Queen's University Belfast, the Harvard-Smithsonian Center for Astrophysics, the Las Cumbres Observatory Global Telescope Network Incorporated, the National Central University of Taiwan, the Space Telescope Science Institute, the National Aeronautics and Space Administration under Grant No. NNX08AR22G issued through the Planetary Science Division of the NASA Science Mission Directorate, the National Science Foundation Grant No. AST-1238877, the University of Maryland, Eotvos Lorand University (ELTE), the Los Alamos National Laboratory, and the Gordon and Betty Moore Foundation.
 This work was partly done using GNU Astronomy Utilities (Gnuastro, ascl.net/1801.009) version 0.17. Work on Gnuastro has been funded by the Japanese Ministry of Education, Culture, Sports, Science, and Technology (MEXT) scholarship and its Grant-in-Aid for Scientific Research (21244012, 24253003), the European Research Council (ERC) advanced grant 339659-MUSICOS, the Spanish Ministry of Economy and Competitiveness (MINECO, grant number AYA2016-76219-P and PID2021-124918NA-C43) and the NextGenerationEU grant through the Recovery and Resilience Facility project ICTS-MRR-2021-03-CEFCA.
The national facility capability for SkyMapper has been funded through ARC LIEF grant LE130100104 from the Australian Research Council, awarded to the University of Sydney, the Australian National University, Swinburne University of Technology, the University of Queensland, the University of Western Australia, the University of Melbourne, Curtin University of Technology, Monash University and the Australian Astronomical Observatory. SkyMapper is owned and operated by The Australian National University's Research School of Astronomy and Astrophysics. The survey data were processed and provided by the SkyMapper Team at ANU. The SkyMapper node of the All-Sky Virtual Observatory (ASVO) is hosted at the National Computational Infrastructure (NCI). Development and support of the SkyMapper node of the ASVO has been funded in part by Astronomy Australia Limited (AAL) and the Australian Government through the Commonwealth's Education Investment Fund (EIF) and National Collaborative Research Infrastructure Strategy (NCRIS), particularly the National eResearch Collaboration Tools and Resources (NeCTAR) and the Australian National Data Service Projects (ANDS).
\end{acknowledgements}
This research has made use of the SVO Filter Profile Service (\url{http://svo2.cab.inta-csic.es/theory/fps/}) supported from the Spanish MINECO through grant AYA2017-84089
%
%
\bibliographystyle{aa}
\bibliography{stein.bib}
\newpage

\begin{appendix}

\section{RECOMMENDATIONS FOR OBSERVING EXTENDED LOW-SURFACE  BRIGHTNESS STRUCTURES WITH AMATEUR TELESCOPES}\label{appendix}

The surface-brightness limits we need to reach reliably for our work on ground is critically affected by numerous technical constraints. This ranges from hardware choices, observing strategies, target selection to calibration procedures and a proper understanding thereof. In the following we provide a summary of recommendations that must be taken into account for future work. Like this, we (i) lower the risk of false detections, (ii) reduce the introduction of artificial features into the background, (iii) keep the background level low, and (iv) allow for better removal of instrumental and environmental fingerprints from the data.

\subsection{Hardware setup}
Straylight supression is mandatory. Direct paths from artificial light sources to the telescope and instrument must be blocked. Light-emitting diodes from local electronics in the observatory dome must be taped to minimise straylight. Heating elements should be used to prevent condensation on sky-facing optical elements that may cool down below ambient temperature. The largest possible filters and adaptor rings etc should be used, to minimize vignetting of the optical path. Any vignetting needs to be corrected afterwards, and the correction by flat fields is never perfect for numerous reasons; chromaticity of the system's spectral response, different spectral energy distribution (SED) of the flat-field source and the night sky, internal scattering are some reasons, to name a few.

\subsection{Choosing suitable observing times}
Strictly dark conditions are mandatory, that is the moon must be below the horizon and the sun at least 18 degrees below horizon. Using a filter that cuts off wavelengths above about 700\,nm will suppress OH airglow lines, which are variable on time scales of minutes and angular scales of arcminutes.

If the sky background is increased by a factor $n$, then the exposure time must be at least $n$ times as long to achieve the same signal-to-noise ratio (SNR), assuming perfectly calibrated and shot-noise-limited data. Reality is worse, because higher sky levels typically means higher systematics that can be hard to remove, degrading the achievable surface-brightness limit. 

Observations in medium- to high humidity conditions must be avoided, as external optical surfaces easily cool below the dew point. Even thin layers of condensation that might not be visible by eye will considerably increase the straylight from in-field and out-of-field sources, reduce the performance of optical coatings, and cause gradients in the image background. This compromises the science data as well as flat-field calibrations. Temperature sensors at the telescope and control of ambient conditions (dew-point temperature) are important. 

Observations must only be conducted in clear conditions. Thin cirrus clouds are not visible to the eye in dark conditions, and all-sky cameras must be used to judge transparency. Guide counts can also be used to evaluate transparency, but they are typically impacted by seeing. Data processing can easily detect transmission variations at the level of 1\% or below, so that bad exposures can be rejected.

\subsection{Target selection}
The FOV must be matched to the target. The data presented in this paper show that stellar streams easily have angular extents exceeding the diameters of their host galaxies by a factor of three or more. The camera's linear FOV should therefore be at least 10 times as large as the major axis of the galaxy. Otherwise one cannot use  sufficiently wide dithers while at the same time keeping the target in the less problematic, unvignetted detector areas.

Airmass is also critical, and targets must be observed as high as possible. If a target is visible at airmass 1.0, then one should not exceed an airmass of 1.3, where the sky background about 30\% higher, and transparency reduced. If a target does not get higher than airmass 1.2, once could plausibly observe it down to 1.4. Targets at even lower airmass are better observed from a different geographic location. 

Another reason to stick to low airmass is the wide FoV, which captures background gradients from differential airmasses. For example, the airmass at $45^\circ$ elevation is 1.41. A FOV of $2^\circ$ would capture airmasses of 1.38 and 1.44 in the same exposure, which could lead to a 4\% variation in background signal that needs to be corrected accurately. 

\subsection{Observing and dithering strategy}

The exposure times must be chosen such that the images are background-limited. That is the photon noise from the sky background must exceed the combined readout and dark-current noise by a factor of five at least. 

The first thing to consider is the maximum expected size of your
target. As the images through out this paper show, at low surface
brightness levels, the maximum "size" of a galaxy (where it becomes
indistinguisable from the noise and undetectable at your final deep
stacked image) is much larger than their brighter regions which are
visible in public sky navigators.

Having determined the exposure time and expected size of your target,
it is important to decide where your target will be placed over each
exposure. Commercial CCDs/detectors are not always a square but can be
rectangular. If your target is not a perfect circle, be sure to
exploit this fact and align the longer axis of your detector to the
major axis of your target.

The most important thing in designing a good observing strategy is to
place your target on very different pixels/regions of your
detector. For example, if your object's maximum length is roughly half
the larger axis of your detector you can do this: in one exposure,
place the edges of your target (accounting for its maximum expected
extent) on the top-left quadrant of your detector. In the next
exposure, place it on the bottom-right, then top-right and finally
bottom-left. Finally, in the fifth exposure, your target can be in the
middle of your frame. If you need more than 5 exposures (which is
usually the case), you can repeat this pointing pattern, but do not
point to the exact same point, allow some random dither (on the scale
of a couple of pixels, but never an exact multiple of the pixel
size).

The yellow regions of the last panel in Figure 1 of \cite{2023RNAAS...7..211A} 
show the deep region after stacking the exposures of the 5-point strategy above. For
a more complete description of this scenario with implementation
details (including how to account for strong vignetting, large areas
of bad pixels, the curvature of the sky and etc), see the dedicated
tutorial within Gnuastro\footnote{\url{https://www.gnu.org/software/gnuastro/manual/html_node/Pointing-pattern-design.html}}. An full implementation of this concept
(to move your target across the your field of view) in a scientific paper can be seen in \cite{2021A&A...654A..40T} (compare the exposure maps
of Figure 1 with Figure 2).

As the two example exposure maps above show, the final result of
stacking/coadding images with this strategy will be larger than your
original detector size, but only the central part will be at full
depth (as you go to the outer regions, the depth decreases
sharply). Because a large fraction of the area of each exposure is not
actually covered by your large and extended target, it can be used for
accurate measurement of the background flux and it will also be
possible to derive the flat-field pattern directly from the science
images (see \cite{2021A&A...654A..40T}). Therefore, the regions that are ultimately shallower
in the final stack have helped in better calibrate the final output.

The 5-pointing example above assumed a ring with a radius that was a
quarter of the longer axis of your detector. If your target is
sufficiently smaller than the limit above, you can use multile rings
with different radii. For example, you can use an inner ring with a
radius of one-eigth of your detector's longer axis. In any case, you
can use the tool described in \cite{2023RNAAS...7..211A} to simulate 
the final stacked image of any pointing pattern before doing the observations 
to make sure they satisfy your plans.

Wide whole sky surveys that use large professional telescopes with
cameras that are composed of many detectors cannot afford to tailor
their observing strategy for individual galaxies like the scenario
proposed here. Their science-cases are also focused more on compact
sources like stars or high redshift galaxies. As a result, the
algorithms they use are are not able to accurately calibrate some
steps (mostly the sky, but also the flat). Therefore dedicated data
taken with commercial telescopes and detectors which are accurately
reduced can be used very well in the study of the low surface
brightness universe.

\subsection{Calibration strategy}
Achieving a uniform background in a flat-fielded image is very difficult, as many factors come into play: vignetting, differential quantum efficiency, different spectrum of the flat-field light source and the night sky background,  external and internal scattered light, internal reflections between filters and collimators, variable plate scale, tree rings, detector doping concentration variations, etc. A simple division by a flat-field cannot correctly account  for all these effects, and is not sufficient for the detection of low-surface brightness features. The dithering strategy outlined above is crucial in this respect, as it allows the computation of suitable correction images that flatten the background (but not necessarily photometric zeropoint variations).

Twilight flats are preferred over dome or screen flats, maintaining the focus of the imaging setup. The flats should be either actively dithered, or the telescope tracking be turned off. The camera angle must be the same as during the night-time observations. It is sufficient to take 10--20 flats per exposure setup. Cameras with even-illumination shutters are preferred, otherwise sufficiently long integration times must be chosen for the flat-fields. 

At least 20 biases should be taken per night, and the bias must be stable within a night. A similar number of dark frames is recommended, and the detector temperature and exposure time must be the same as for the science exposures. Flat fields usually do not require matching dark frames, biases are sufficient.

\subsection{Detector considerations}
Large detectors can have multiple, configurable readout ports. A single readout port should be used, so that all pixels experience the same systematics by the electronics chain. Detector systems and cameras that do not allow full control, that adjust internal settings on the fly or even apply some level of pixel processing, should be avoided.

\end{appendix}
\end{document}